\documentclass[
superscriptaddress,
 amsmath,amssymb,
 aps,
 prl,
 twocolumn
]{revtex4-2}

\usepackage{float}
\usepackage{graphicx}
\usepackage{braket}
\usepackage{amsmath}
\usepackage{inputenc}
\usepackage{stix}
\usepackage[dvipsnames]{xcolor}
\usepackage[colorlinks,linkcolor=blue,citecolor=blue,urlcolor=blue]{hyperref}

\renewcommand{\vec}[1]{\mathbf{#1}}

\begin{document}

\preprint{APS/123-QED}

\title{Correlation-Assisted Quantized Charge Pumping}

\author{Jacob A. Marks}
\email{jamarks@stanford.edu}
 \affiliation{Physics Department, Stanford University, Stanford, CA 94035, USA}
 \affiliation{Stanford Institute for Materials and Energy Sciences (SIMES),
	SLAC National Accelerator Laboratory, Menlo Park, CA 94025, USA}
\author{Michael Sch\"uler}%
 \email{schuelem@stanford.edu}
 \affiliation{Stanford Institute for Materials and Energy Sciences (SIMES),
SLAC National Accelerator Laboratory, Menlo Park, CA 94025, USA}%

\author{Jan C. Budich}
\affiliation{Institute of Theoretical Physics, Technische Universit\"{a}t Dresden and W\"urzburg-Dresden Cluster of Excellence ct.qmat, 01062 Dresden, Germany}

\author{Thomas P. Devereaux}
\affiliation{
 Department of Materials Science and Engineering, Stanford University, Stanford, CA 94035, USA}
\affiliation{Stanford Institute for Materials and Energy Sciences (SIMES),
SLAC National Accelerator Laboratory, Menlo Park, CA 94025, USA}

\date{\today}

\begin{abstract}
We investigate charge pumping in the vicinity of order-obstructed topological phases, i.e. symmetry protected topological phases masked by spontaneous symmetry breaking in the presence of strong correlations. To explore this, we study a prototypical Su-Schrieffer-Heeger model with finite-range interaction that gives rise to orbital charge density wave order, and characterize the impact of this order on the model's topological properties. In the ordered phase, where the many-body topological invariant loses quantization, we find that not only is quantized charge pumping still possible, but it is even assisted by the collective nature of the orbital charge density wave order. Remarkably, we show that the Thouless pump scenario may be used to uncover the underlying topology of order-obstructed phases.
\end{abstract}

\maketitle

The robust quantization of transport properties observed in topological states of quantum matter is among the most fascinating phenomena in physics~\cite{hasan_colloquium:_2010, qi_rmp_2011}, both from a fundamental perspective and due to its far-ranging potential for technological applications~\cite{chang_experimental_2013, konig_quantum_2007, yue_nanometric_2017}. A primary example along these lines is provided by the integer quantum Hall effect~\cite{klitzing_prl_1980, prange_qhe_1987} in two-dimensional systems with a finite Chern number~\cite{tknn_1982}, as
induced by a strong perpendicular magnetic field. Seminal work by Laughlin~\cite{laughlin_1981} and Thouless~\cite{thouless_prb_1983} has revealed that this topologically quantized charge transport may be understood as a cyclic adiabatic pumping process in time-dependent
one-dimensional systems. While these intriguing phenomena can be understood within the independent particle approximation in the framework of topological Bloch bands~\cite{qi_rmp_2011}, their stability against imperfections such as disorder and weak to moderate correlations is well established~\cite{haldane_model_1988,hasan_colloquium:_2010,qi_topological_2008}.  In strongly correlated systems however, qualitative changes to this picture occur, including the breakdown of topological band theory due to spontaneous symmetry breaking~\cite{Rachel_2010, Hohenadler_prl_2011} and dynamical quantum
fluctuations~\cite{budich_fluctuation-induced_2012, kka_prl_2014, Amaricci_prl_2015}, respectively, but also the formation of
genuinely correlated topologically ordered phases~\cite{wen_rmp_2017}.

\begin{figure}[t!]
\includegraphics[width=0.45\textwidth]{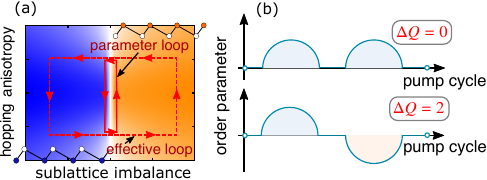}%
\caption{\label{fig:cycles_sketch} Schematic of correlation-assisted charge pumping. (a): Sketch of correlation-assisted pump cycle in parameter space (solid line) and the effective parameter cycle in the absence of correlations (dashed line). Blue and red background in phase diagram denote orbital character of sublattice ordering (sublattice A and B respectively). (b): Sketch of order parameter over two correlation-assisted pump cycles which do not (upper) and do (lower) pump non-zero charge. Both cycles enter and exit the ordered phase, but only in the bottom cycle, corresponding to the cycle in (a), does the ordering change orbital character, leading to non-zero pumped charge.}
\end{figure}

In this work, we demonstrate how spontaneous symmetry breaking can facilitate and even drive quantized charge pumping (see sketch in Fig.~\ref{fig:cycles_sketch}). Remarkably, by means of unbiased numerical simulations, we show that this mechanism survives even in a strongly correlated regime, where an effective single-particle picture is found to break down, and would not correctly predict the adiabatic pumping properties. In particular, we observe that the Resta polarization~\cite{resta_quantum-mechanical_1998} provides an unambiguous many-body topological characterization that agrees with direct calculation of the relevant transport properties. Furthermore, we reveal how the Thouless charge pumping approach~\cite{Niu_1984, thouless_prb_1983} can be used to characterize the buried topological phase diagram of order-obstructed phases, i.e. conventionally ordered phases that arise from symmetry protected topological (SPT) phases~\cite{chen_spt_prb_2013} by spontaneous symmetry breaking. We also note related recent works on the interplay between topology and symmetry breaking~\cite{cuadra_prb_2019, cuadra_natcomm_2019, cuadra_soliton_2019}, as well as other work on quantized charge pumping~\cite{PhysRevB.91.125411, PhysRevB.96.085444}, and on topology in strongly correlated systems~\cite{Hetenyi_2020}.

To this end, we study in the framework of density matrix renormalization group (DMRG) methods~\cite{white_dmrg_1992, schollwock_density-matrix_2005} a variant of the Su-Schrieffer-Heeger
(SSH) model~\cite{su_solitons_1979, heeger_solitons_1988} with finite-range interaction as a conceptually simple system exhibiting rich interplay between SPT phases and long-range order. For strong interactions, our model system exhibits an orbital charge-density wave (CDW) which
spontaneously breaks the protecting chiral symmetry, exemplifying the aforementioned class of order-obstructed phases. Our findings closely connect to current experimental activity on realizing topological band structures with ultracold atoms in optical lattices~\cite{aidelsburger_experimental_2011, struck_tunable_2012, goldman_topological_2016, dauphin_loading_2017, flaschner_experimental_2016}, and may be verified in a strongly interacting version of recent experiments on quantized charge pumping in such settings~\cite{lohse_thouless_2016, nakajima_topological_2016}.

\paragraph{Model and Methods.}---
Initially conceived as a microscopic description of solitons in polyacetylene~\cite{heeger_solitons_1988}, the SSH model has become a prototype for topological physics in one-dimensional systems. Several interacting variants of the SSH model have been studied in previous literature~\cite{junemann_exploring_2017, sbierski_topological_2018, marques_multihole_2017}, where extended interactions have been seen to give rise to exotic collective phases of matter~\cite{anisimovas_semisynthetic_2016}. Here, considering a extended interaction $\hat{V}$ that retains both chiral and particle hole symmetry ~\cite{kruckenhauser_dynamical_2018, thermalizing}, we demonstrate how the interplay of SPT order and spontaneous symmetry breaking conspire to exhibit interesting, previously undocumented behavior. We note that a similar model is considered in~\cite{Yahyavi_2018}.

The particular model we study is described by the microscopic Hamiltonian:
\begin{align} \label{eq:model}
\hat{H}& =- \sum_j \left( J \hat{b}^{\dag}_j \hat{a}_j +  \frac{d + \tau}{2} \hat{b}^{\dag}_j \hat{a}_{j+1} +  \frac{d - \tau}{2} \hat{a}^{\dag}_j \hat{b}_{j+1}\right) + \mathrm{h.\,c.}\nonumber \\ 
& + \quad V \sum_j \left(\hat{n}^a_j \hat{n}^b_{j + 1} + \hat{n}^b_j \hat{n}^a_{j + 1} \right),
\end{align}
on a bipartite chain of spinless fermions with two orbitals per site ($\mathrm{a}$ and $\mathrm{b}$). $\hat{a}^{\dag}_j (\hat{a}_j)$ and $\hat{b}^{\dag}_j (\hat{b}_j)$ represent creation (annihilation) operators on orbitals in sublattice $a$ and $b$ respectively, on site $j$. $J$ is an intra-cell hopping~\footnote{Our analysis strictly excludes the point $J = 0$, at which the model obeys different symmetries.}, and $(d\pm \tau)/2$ are alternating inter-cell inter-orbital hopping strengths. In the second summation, $\hat{V}$, $\hat{n}^{a} = \hat{a}^{\dag} \hat{a}$, and likewise for $\hat{n}^{b}$. To investigate this model, we use the DMRG technique~\cite{schollwock_density-matrix_2005}. Below, we show results for periodic boundary conditions (PBC). In the Supplementary Material~\cite{supplement} (see also references~\cite{bardyn_probing_2018, fu_kane, altland_nonstandard_1997} therein), we present analogous results for systems with open boundary conditions (OBC) and extensively compare the two. For DMRG and system details, see~\cite{supplement}.

To probe topological properties of our model (\ref{eq:model}), we focus on two many-body topological invariants. First, the Resta (many-body) polarization for periodic systems~\cite{resta_quantum-mechanical_1998},
\begin{align}
P = \frac{q a_0}{2 \pi} \mathrm{Im} \ln \mathrm{Tr}\left[\hat{\rho}\, e^{\frac{i 2 \pi}{L a_0} \hat{X}}\right] \quad (\mathrm{mod} \,q a_0) \ ,
\label{eq:restapol}
\end{align}
where $a_0$ is the lattice spacing, $L$ is the number of unit cells, and $\hat{X} = \sum_i x_i \hat{n}_i$ is the many-body center of mass operator; $\hat{\rho}$ denotes the many-body density matrix~\footnote{$P$ defined in this way is only strictly valid as a topological invariant for periodic systems. The polarization for OBC systems  is defined as $P_{OBC} = q\langle \hat{X}\rangle$. However, $P$ is still well-defined for
OBC systems as long as the effects of single particle edge modes
are restricted to the boundaries.}. Second, we study the entanglement spectrum~\cite{pollmann_entanglement_2010, fidkowski_topological_2011, li_entanglement_2008}, which is intimately connected to the Resta polarization~\cite{Zaletel_2014}, and to charge pumping~\cite{hayward_topological_2018}. It has also been studied specifically in the context of the SSH model~\cite{Ryu_Hatsugai_2006}. To define this, we consider the reduced density matrix resulting from tracing out half of our system across a spatial bipartition: $\rho_L = \mathrm{Tr}_R [\rho_{LR}]$. The entanglement spectrum is the set of ordered eigenvalues $\lambda_i$ of the entanglement Hamiltonian $\mathcal{H}_E = - \mathrm{log}\rho_L$. The degeneracies of the spectrum are determined by fundamental characteristics of the state, including topological properties and symmetries such as inversion symmetry. We define the \textit{entanglement gap} as $\Delta \lambda = \lambda_1 - \lambda_0$.

\begin{figure}[t!]
\includegraphics[width=0.45\textwidth]{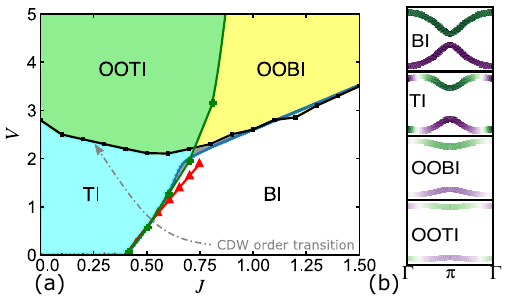}%
\caption{\label{fig:phase_diagram}(a): Phase diagram for interacting SSH model with $d = 0.4$, $\tau = 1.0$. \textbf{TI}: Topological Insulator, \textbf{BI}: Band Insulator, \textbf{OOBI}: Order-obstructed Band Insulator, \textbf{OOTI}: Order-obstructed Topological Insulator. Red triangles denote mean field theory (MFT) results for \textbf{BI} $\rightarrow$ \textbf{TI} phase boundary. Black squares denote finite size scaling calculations from fitting the orbital CDW phase transition to Ising universality class. The blue line delineates regions of single (right) and double (left) ground state degeneracy for OBC systems. Green '$+$' and solid green line indicates polarization phase boundary. Simulations were performed on systems with up to $L = 100$ unit cells. Gray shading in center represents region of uncertainty given the accessible system sizes. (b): Typical band structures in each phase. 
Color signifies orbital character with respect to the BI basis, and level of transparency represents strength of band mixing.}
\end{figure}

\paragraph{Phase Diagram.}---
For concreteness, we focus on the phase diagram slice for $d = 0.4, \tau = 1.0$  (cf. Fig.~\ref{fig:phase_diagram}~(a)). However, we investigated many parameter combinations and found that the qualitative behavior is generally preserved across the parameter regimes considered. Roughly, increasing $d$ shifts the trivial-topological phase boundary to the right, and $\tau$ does not have much impact on the topological properties (as long as $\tau \neq 0$ ). Our designations of trivial and topological regions in the phase diagram are determined by the value of the polarization (\ref{eq:restapol}) in the normal phase (defined by explicitly suppressing CDW order). This is consistent with our complementary classification approach by entanglement. CDW order can be identified by the change in ground state degeneracy~\footnote{This is true for PBC systems, where only the OOTI and OOBI phases have doubly degenerate ground states. For OBC systems the TI phase does as well, as illustrated by the blue line in \ref{fig:phase_diagram}~(a). More sophisticated measures must be taken for OBC systems, as detailed in the Supplementary Material~\cite{supplement}}.

In the resulting phase diagram of the SSH model with interaction (\ref{eq:model}), we identify four different phases, distinguished by the presence or absence of SPT order and CDW order (see Fig.~\ref{fig:phase_diagram}~(a)). The first two phases, band insulator (BI) and topological insulator (TI) are present without interaction and persist for weak to moderate interaction. For sufficiently strong interaction (indicated by the black squares in Fig.~\ref{fig:phase_diagram}~(a)), spontaneous orbital CDW order emerges (see~\cite{supplement}), which obstructs the underlying topology. The two resulting phases, OOBI and OOTI, are characterized by sublattice symmetry breaking. Their topological properties are discussed below. We checked that the phase diagram is not qualitatively changed by the addition of other small terms obeying chiral symmetry, or by the presence of a small intra-cell interaction $\hat{U}$. Subsequently, we elaborate on all four phases, going through the phase diagram in Fig.~\ref{fig:phase_diagram}~(a) from weak to strong interactions.

For $V = 0$, the Hamiltonian can be expressed in the single-particle basis, and all topological information can also be extracted from the single particle density matrix (SPDM) by decomposing
$\rho_k(t) = \frac{1}{2} \big[ 1 - \vec{r}_k(t) \cdot \vec{\sigma} \big]$ in momentum-space, where $\vec{r}_k$ is the pseudospin vector. In equilibrium, the quantity $\nu$, defined by $(-1)^{\nu} = \mathrm{sgn}(r^x_{\Gamma} \, r^x_{\pi})$, coincides with the pseudospin winding number~\cite{perez-gonzalez_ssh_2018}:  $\mathcal{W} = \frac{1}{2 \pi} \int^{\pi/a}_{-\pi/a} dk (n^x_k \partial_k n^y_k - n^y_k \partial_k n^x_k)$, where $\mathbf{n}_k = \mathbf{r}_k/|\mathbf{r}_k|$. A nontrivial value of either invariant is equivalent to the condition $|d| > |J|$, and a topological phase transition occurs when $|\vec{r}_{\Gamma}| = 0$, at which point $r^x_{\Gamma}$ changes sign. One can also view the SPDM as the density matrix corresponding to the (in the presence of interactions, mixed) state of some auxiliary Hamiltonian, $\rho_k = e^{- \mathcal{H}_k}$
\footnote{In equilibrium, $\mathcal{H}_k$ is proportional to the single-particle Hamiltonian.}, and can use $\mathcal{H}_k$ to compute the Zak phase~\cite{zak_berrys_1989} from its eigenstates: $\mathcal{Z} = \frac{i}{\pi} \int^{\pi/a}_{-\pi/a} dk \langle \phi_k | \partial_k \phi_k \rangle$. The Zak phase is quantized to $\mathcal{Z}=\nu$ and can be used as a topological invariant in the non-interacting SSH model; it is also equivalent to the quantized polarization, $P=\nu q a_0/2$. 

When weak interaction (preserving chiral symmetry) is added to the model, the Zak phase and the topological invariant extracted from the SPDM both agree with the Resta polarization. This reflects the fact that the SPDM still contains most of the information about the many-body topology~\cite{supplement}. The positive slope of the BI $\rightarrow$ TI phase boundary can be attributed to (even at mean-field level) the interaction pushing valence and conduction bands closer together, facilitating hybridization. As illustrated in the band-structure sketches in Fig.~\ref{fig:phase_diagram}~(b), the TI phase is characterized by band inversion at the $\Gamma$-point.

\begin{figure}[b!]
\includegraphics[width=0.48\textwidth]{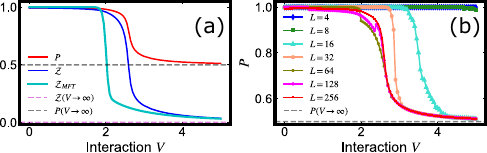}%
\caption{\label{fig:fractionalization}Fractionalization in the symmetry broken phase, with $J = 0.1$, $d = 0.4$, $\tau = 1.0$. (a) Resta polarization $P$ computed using DMRG, and the Zak phase, computed from DMRG and MFT for $L = 256$ unit cells. In infinite $V$ limit in the symmetry broken phase, $\mathcal{Z} \rightarrow 0$ and $P \rightarrow \pm q a_0/4$. (b) System-size dependence of the polarization. Polarization curve converges with system-size, but does not obey finite-size scaling. $P$ is in units of $qa_0/2$.}
\end{figure}

Strong interaction favors the emergence of orbital CDW order with ground states that spontaneously break chiral symmetry. When explicitly suppressing CDW order (thus enforcing chiral symmetry), the solid green line in Fig.~\ref{fig:phase_diagram}~(a) still separates the topological insulator from the trivial insulator phase. In this scenario, we can distinguish these two phases by their polarization ($P=0$ for trivial, $P=\pm q a_0/2$ for topological) even within the unstable region where in principle charge order would obstruct the SPT phases. Furthermore, the strongly interacting topological region still features edge modes. When allowing for CDW order, the topologically trivial phase defines the OOBI region in Fig.~\ref{fig:phase_diagram}~(a) within which no charge can be pumped, and the topological phase defines the OOTI region in Fig.~\ref{fig:phase_diagram}~(a), within which quantized charge pumping is possible as shown below.

Nevertheless, even in the order obstructed regime, important signatures of the inherited topological properties persist, as illustrated in Fig.~\ref{fig:phase_diagram}~(b). The trivial OOBI has a band structure similar to BI, but with strong band mixing close to the $\Gamma$-point. The band structure of OOTI combines features of TI and OOBI: band inversion and band mixing. The onset of orbital order for transitions TI $\rightarrow$ OOTI and BI $\rightarrow$ OOBI can be identified by entanglement signatures in the symmetry broken state~\cite{supplement}. In the band insulating phase, the transition is accompanied by an abrupt change in the slope of the entanglement entropy. In the TI phase, the entanglement gap is $\Delta \lambda = 0$, and becomes non-zero upon emergence of orbital order. In the OOTI phase, the edge states are gapped out. Remarkably, even in the presence of CDW order, we will demonstrate that adiabatic charge pumping still allows us to distinguish the OOTI from the OOBI region, as separated by the solid green line in Fig.~\ref{fig:phase_diagram}~(a).

To compare the single-particle and the many-body characterization of the topology, we compute the natural orbitals $|\phi_k\rangle$ from the SPDM as the best choice of a single-particle basis and the corresponding Zak phase $\mathcal{Z}$. The breakdown of such single-particle topological invariants was previously documented~\cite{grusdt_topological_2019-1}.
Remarkably, the Resta polarization $P$ and $\mathcal{Z}$ agree well for large parts of the phase diagram. In the normal phase, both  $\mathcal{Z}$ and $P$ remain quantized, and the two quantities align for all $V$~\footnote{For OBC systems, $P$ is well-behaved in the symmetry broken phase (in qualitative agreement with PBC), in the normal phase experiences a boundary-driven transition for moderate system sizes. This (finite-size effect) transition, which is due to edge states hybridizing with the bulk, disappears in the thermodynamic limit.}. In the ordered phase, both $P$ and $\mathcal{Z}$ fractionalize (Fig.~\ref{fig:fractionalization}~(a)). In the large $V$ limit, hallmarking the breakdown of the single-particle picture, the polarization of the degenerate ground states approaches the fractionalized values $P = \pm q a_0/4$~\footnote{The fractionalized value for $P$ is intrinsically due to CDW order.}, and the Zak phase converges towards $\mathcal{Z} = 0$. Fig.~\ref{fig:fractionalization}~(b) illustrates the system-size dependence of the $P$. While $P$ (as a function of $V$) converges with system-size, it inherits only indirectly from the CDW order, and as such does not obey the same scaling collapse.

\paragraph{Charge Pumping.}---
Despite the fact that polarization (\ref{eq:restapol}) is not quantized in the ordered phase, we find that the underlying topological character of the phase without broken symmetry is present in the adiabatic transport properties of the system. One can imagine adding a (infintesimally small) staggered on-site potential term $\hat{\Delta} = \Delta \sum_j (\hat{n}^a_j - \hat{n}^b_j)$ to the model, which acts as a pinning field. The point $\Delta = \tau = 0$ is the degeneracy point, at which the system has no preference for either sub-orbitals $\mathrm{a}$ or $\mathrm{b}$. Adiabatically looping around this degeneracy point through cyclic variation of $\Delta$ and $\tau$, quantized charge 
\begin{align}
    \Delta Q = \frac{1}{a} \int^{T}_{0} dt \, \partial_t P(t),
\end{align}
is transported. In practice, we compute $P(t)$ along the cycle from the instantaneous Hamiltonian $\hat{H}(\theta)$ with the loop variable $\theta \in [0,2\pi]$~\footnote{We have also performed these cycles explicitly via time evolution on small systems. The results are presented in the Supplementary Material~\cite{supplement}}.

\begin{figure}[t!]
\includegraphics[width=0.5\textwidth]{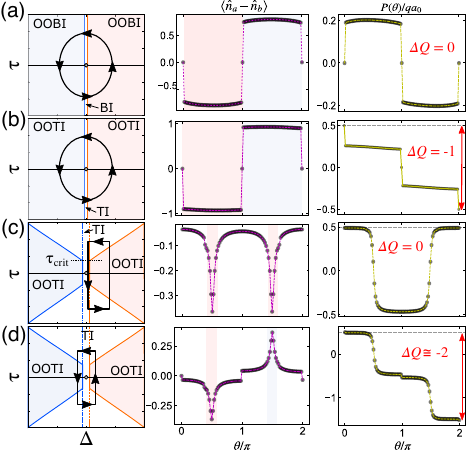}%
\caption{\label{fig:pumps} Correlation-assisted Thouless pump cycles. Left panel: sketch of cycle in phase diagram. Center panel: order parameter along cycle. Right panel: polarization along cycle. (a) and (b): Pump cycles within (a) OOBI and (b) OOTI phases. In (c) and (d), the system enters and exits the ordered phase. (d) successfully transports charge, but (c) does not. Shaded blue (orange) regions denote orbital ordering with occupation localized on sublattice $\mathrm{a}$ ($\mathrm{b}$), across which the order parameter attains peak magnitude in the thermodynamic limit. In numerics, (d) transports charge $\Delta Q = -2.010(4)$, which is asymptotically quantized to $\Delta Q \rightarrow 2$ in the thermodynamic limit. Each cycle performed by adiabatically connecting $100$ independent ground state calculations, for systems of $L = 100$ unit cells. Specific parameter values used to generate these results are detailed in Table~I in the Supplementary material~\cite{supplement}.}
\end{figure}

Inspecting $\Delta Q$ across the phase diagram Fig.~\ref{fig:phase_diagram}~(a), we find that the topological character of the underlying state \emph{without} broken symmetry is recovered. While this behavior is expected for the weakly-interacting TI phase, it is remarkable that the obstructed OOTI phase (with $P\ne 0$ without broken symmetry) is characterized by quantized $\Delta Q\ne 0$. This is demonstrated in Fig.~\ref{fig:pumps}~(a)--(d).

The unit of quantization in OOTI is $1/2$ that of the TI phase, reflecting the fractionalization of excitations in the presence of orbital order. Additionally, while charge is pumped continuously in TI, in OOTI charge is mostly pumped in discrete jumps, \emph{driven} by the collective order.

These jumps are easily understood in the context of sublattice filling: The cycle begins with the system in the insulating phase. Very abruptly the onset of order leads to orbital occupation supported primarily on sublattice $\mathrm{a}$. This `jump' is accomplished by shifting half of the electron occupation (the occupation of sublattice $\mathrm{b}$ orbitals) to the a orbital on the same site. When $\theta = \pi$, $\Delta$ changes sign, quickly altering the energy landscape to favor occupation on $\mathrm{b}$ rather than $\mathrm{a}$. In this moment, the second jump occurs as all of the electrons shift to the neighboring sublattice $\mathrm{b}$. Finally, the system re-enters the topological insulating phase and half of the occupation moves from $\mathrm{b}$ to the $\mathrm{a}$ orbital on the same site. Altogether, one unit of charge is pumped during this cycle. Also note that the jump at $\theta = \pi$ is twice as large as the other jumps, as \emph{all} rather than half of the occupation shifts one orbital.

Moreover, the single-particle picture of charge pumping breaks down in the presence of strong interactions. An effective single-particle description is obtained from integrating the Berry curvature of the natural orbitals $\Omega(k,t)=2 \, \mathrm{Im}\langle \partial_k \phi_k(t) | \partial_t \phi_k(t) \rangle$: $\Delta Q = (q/2\pi) \int\!dk \int\!dt \Omega(k,t) n(k,t)$, where $n(k,t)$ is the larger occupation eigenvalue. Similar to Ref.~\cite{hayward_topological_2018}, we find that the non-uniform $n(k,t)$ along the cycle leads to non-quantized $\Delta Q$ in this single-particle picture.

Whereas the pump cycles in Fig.~\ref{fig:pumps}~(a) and (b) demonstrate that charge pumping remains quantized in the order-obstructed phase, Fig.~\ref{fig:pumps}~(c) and (d) illustrate how charge pumping is facilitated by the collective nature of the obstructing order. For these two cycles, there is a critical $\tau$ that controls whether or not there is orbital ordering when an infinitesimal seed is added. As such, these cycles utilize the spontaneous symmetry breaking of the ground state to minimize reliance on on-site staggered potential $\Delta$. In the ordered phase, orbital character of the state is completely determined by the sign of $\Delta$, independent of magnitude. Thus, one can exert control over the orbital character (for $\tau$ in the right region) via infinitesimal staggered onsite potential (or any other mechanism which acts as a seed for the order). In (c), both times $\tau$ becomes small enough to induce ordering it settles on the same orbital character, resulting in net zero charge pumped. In (d) however, $\Delta$ changes sign, leading to four jumps in sublattice occupation (or two pairs of jumps).

As $|\tau| \rightarrow 0$, the system spontaneously orders before reaching the degeneracy point, avoiding the band-gap closing~\footnote{Due to finite-size effects, we were forced to used small pinning field $|\Delta| = 10^{-2}$ throughout. In the thermodynamic limit however, $\Delta = 0$ would also work (the collective order can be induced by random fluctuations).}. Since this spontaneous ordering randomly picks a direction, not all pump cycles will transport net charge, as demonstrated in Fig.~\ref{fig:pumps}~(c). If the sign of the order coincides with that of $\tau$ as in Fig.~\ref{fig:pumps}~(d), then charge is indeed transported. This illustrates the possibility of quantized charge transport with neither bias nor staggered potential. 

In this example, the correlation-assistance to charge pumping comes in the form of eliminating the staggered potential, enabling quantized pumping by only varying the hopping anisotropy. In general, it operationally means one can non-trivially control charge pumping via a single tunable parameter, greatly increasing feasibility of experimental efforts~\footnote{These correlations also allow us to realize faster pumps, as explicitly illustrated in~\cite{supplement}}.

\paragraph{Concluding discussion.}---
In summary, we have described and characterized the effect of spontaneous symmetry breaking on many-body topology, and have shown that the Thouless pump can be used to identify underlying topology in order-obstructed phases. Moreover, we illustrate that collective order can be conducive to quantized charge transport, with spontaneous ordering working to prevent band gap closing even when the model is tuned to the degeneracy point. En route to demonstrating this effect, we have presented a complete phase diagram for an SSH model with extended interaction, including a phase in which orbital charge density wave order obstructs topology. Our model may be realized in state of the art experiments on cold-atoms in optical lattices \cite{lohse_thouless_2016}.

The correlation-assisted charge pumping we demonstrate in the interacting SSH model should be typical of the interplay between topology and collective order. In our present case of a spontaneously broken $\mathbb{Z}_2$ symmetry, random fluctuations alone suffice to produce quantized charge transport. For other symmetries, what survives is that only an infinitesimally small external potential is needed to control transport processes.
The collective phase can be exploited to circumvent the many-body topological constraint in real-time evolution, where the topological invariant is pinned to its initial value under unitary evolution. Hence, a dynamical topological phase transition can be induced by passing through the symmetry breaking collective phase, giving rise to dynamically induced symmetry breaking~\cite{mcginley_topology_2018}. This principle can be generalized to ordered phases obstructing other SPT states.

\begin{acknowledgments}
{\textit{Acknowledgements. --} }DMRG calculations were performed using the ITensor library \cite{noauthor_itensor_nodate}. We thank Miles E. Stoudenmire for helpful discussions regarding time evolution of Fermionic systems in ITensor. We also thank the Stanford Research Computing Center for providing computational resources. Data used in this manuscript is stored on Stanford's Sherlock computing cluster. Supported by the U.S. Department of Energy (DOE), Office of Basic Energy Sciences, Division of Materials Sciences and Engineering, under contract DE-AC02-76SF00515. M. S. thanks the Alexander von Humboldt Foundation for its support with a Feodor Lynen scholarship. J.C.B.\ acknowledges financial support from the German Research Foundation (DFG) through the Collaborative Research Centre SFB 1143 (Project No.~247310070) and the W\"urzburg-Dresden Cluster of Excellence on Complexity and Topology in Quantum Matter -- ct.qmat (EXC 2147, Project No.~39085490). 
\end{acknowledgments}


%

\clearpage

\appendix

\renewcommand{\thefigure}{A\arabic{figure}}
\renewcommand{\theequation}{A\arabic{equation}}
\setcounter{figure}{0} 
\setcounter{equation}{0}

\section{APPENDIX}
While results presented in the main text utilize DMRG simulations on systems with periodic boundary conditions (PBC), we emphasize that the physics we have illustrated also holds (up to well-understood finite-size effects) for systems with open boundary conditions (OBC). In these appendices, we support these claims through extensive simulations on OBC systems. We also bolster our PBC results with in-depth discussion of the Resta polarization in the order-obstructed phases, as well as explicit time evolution with time-dependent DMRG (tDMRG) simulations. 

Appendix A provides further details about the model, including a single-particle description in the non-interacting model, and plots of the many-body energy gap for OBC systems.

In Appendix B, we apply Hartree-Fock treatment as a viable description of the weakly-interacting regime, and in the process show that translation-invariance is a suitable approximation for large enough OBC systems. 

In Appendix C, we corroborate our results for the strongly-interacting regime with an effective spin model description, and numerically demonstrate scaling-collapse of the CDW phase transition to the classical $2d$ Ising universality class. 

Appendix D provides technical details of our investigation, including DMRG specifications and optimization routines. 

Appendix E develops intuition and engages with the mathematics underlying the fractionalization of the Resta polarization in the order-obstructed phases.

In Appendix F we illuminate why the Resta polarization, which is only well-defined for periodic systems, serves as a good OBC many-body topological invariant and concurs with that of PBC systems - in all but the very strongly-interacting regime. With this insight, we expressly demonstrate that one can numerically probe the topological character of order-obstructed SPT phases via DMRG simulations on OBC systems, which often offer drastically-reduced computational cost. In particular, we explicitly simulate the pump cycles in Fig.~4 in the main text using OBC (again by adiabatically connecting the polarization of independent ground-state computations through the cycle), and verify that the results are qualitatively - and quantitatively up to negligible finite-size effects - unchanged.

Finally, performing pump cycles explicitly by imaginary time evolution, in Appendix G we corroborate our conclusions regarding correlation-assisted charge pumping.

\begin{figure}[htp!]
\includegraphics[width=0.45\textwidth]{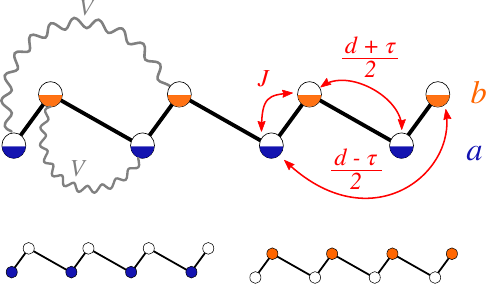} 
\caption{Sketches of the system considered in the main text. Top: system in insulating phase, with half-filling on each sublattice. Bottom: system with orbital ordering of character A (left) and B (right). \label{fig:system_sketch}}
\end{figure}

\begin{figure*}[htp!]
\includegraphics[width=\textwidth]{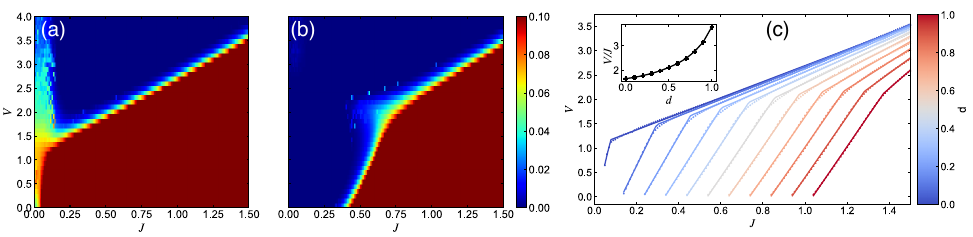} 
\caption{Energy degeneracy computations for OBC systems. (a) and (b) Non-linearly scaled energy difference of lowest lying states, according to $1 - \mathrm{exp}(-\Delta E)$. (a) $d = 0.0$; Yellow region near $J = 0$ due to finite size effects (b) $d = 0.4$; kink in center coincides with crossover from mean-field theory to ordered regime (c) piece-wise linear fit separating singly degenerate (bottom right) from doubly degenerate (upper, left) ground states. Inset: quadratic dependence of slope $V/J$ of  BI-OOBI phase boundary. \label{fig:energy_diffs}}
\end{figure*}

\section{Appendix A: Sketch of System and Phase Diagram Details}
Fig.~\ref{fig:system_sketch} illustrates various phases of the SSH model described by Eq.~(1) in the main text. In the non-interacting case, the SSH model is exactly solvable, making it particularly well-suited to study the effects of interaction on edge states, and providing insight into higher-dimensional topology. In the main text we consider a chain of spinless fermions, as the spin degree of freedom adds unnecessary complexity for the physics we wish to elucidate. However, we note that the SSH model can easily be extended to spinful fermions, and the effects we document survive. Moreover its topology is protected by chiral and time-reversal symmetry~\cite{altland_nonstandard_1997}, meaning that it is a class BDI system in the Altland-Zirnbauer classification of random matrices. Thus, the topology should survive in the presence of disorder or weak interactions that preserve these symmetries. Absent interaction, the SSH Hamiltonian can be written in the single-particle basis as $\hat{H}_0 =- \sum_j \hat{\vec{c}}^{\dag}_k [\vec{h}_k \cdot \vec{\sigma} ] \hat{\vec{c}}_k$, where $\hat{\vec{c}}_k = (\hat{a}_k,\hat{b}_k)$ is a spinor whose components annihilate fermions on sublattice $a$, $b$ respectively with lattice momentum $k$. $\vec{\sigma }$ denotes the vector of Pauli spin matrices, and $\vec{h}_k = \big(-J - d \cos{k}, -\tau \sin{k}, 0 \big)$, where $h^z_k = 0$ due to chiral symmetry. Instead of analyzing the single-particle Hamiltonian, we can analyze the SPDM - this  provides a natural way of treating mixed states, interacting systems and non-equilibrium scenarios~\cite{bardyn_probing_2018}. This simple single-particle description also explains our seemingly complicated choice of hopping terms $J$, and $(d\pm \tau)/2$. When the many-body Hamiltonian is written in this fashion, it becomes clear that $|d| \lessgtr |J|$ determines the topology, and one can vary $\tau$ along a charge pump cycle without affecting the topology. 

We also comment on the inclusion of third neighbor hopping terms and the omission of second neighbor terms. Second-neighbor hopping is strictly forbidden by the symmetries of our system, which protect the topology. This means that third-neighbor hopping is the simplest addition one can make to the model. The introduction of third neighbor hopping terms qualitatively changes the physics. The band structures are significantly altered, and the dispersion relations shed their single minimum and acquire Mexican-hat-like shape. The correlation-assisted charge pumping phenomena described in the main text are robust to the particular coefficients of first and third neighbor hopping - so long as both are non-zero - and persist in all investigated regimes.

We have thoroughly attempted to generate correlation-assisted charge pumps with only nearest-neighbor hopping, and have not found any such parameter combinations that pump non-zero charge in correlation-assisted fashion. This can be explained as follows: in our model, eliminating third-neighbor terms is equivalent to setting $d = \tau$, which ties these two parameters together. In our pump cycles, $\tau$ was used as one of the two parameters varied over the cycle. Each cycle must wind around the degeneracy point, $\Delta = \tau = 0$, and if $\tau$ and $d$ are ties, then $|d| < |J|$ by necessity at some point along the cycle, and the system exits the topological phase.

In the large $V$ limit, there are two ground states. in the thermodynamic limit, one has fully occupied (empty) sublattice $\mathrm{a}$ ($\mathrm{b}$) orbitals. The other ground state is the particle-hole symmetric pair state. Generically, small random fluctuations will drive the system into the symmetry broken state, and the SPT phase is obstructed by spontaneous symmetry breaking. In both the OOBI and OOTI phases, the lowest-order corrections to the ground state are of order $\mathcal{O}\big(J/V, (d \pm \tau)^2/V \big)$, which means that the effect of $J$ survives longer than $d$ or $\tau$, and as $V \rightarrow \infty$, we obtain an effective Ising ferromagnet. Furthermore, the topological character of the system cannot disappear at the onset of the orbital CDW order, as that would imply the existence of a local order parameter for topology.

The breakdown of the single-particle picture can be understood in terms of the relative sizes of the single and many-body gaps. To the left of the green line in the phase diagram Fig.~2~(a), even though the pseudospin gives a winding number of $\mathcal{W} = 1$ for all $V$, the single-particle density matrix becomes increasingly mixed with increasing $V$. This is visualized for OBC systems in Fig.~\ref{fig:pseudospin}, where the radius of the pseudospin covering $\mathbf{r}_k$, (which is still confined to the $xy$~-~plane) goes to zero, while still encircling the origin as $V \rightarrow \infty$. This means that the single particle Hamiltonian remains gapped, but the single-particle gap approaches zero. On the single-particle level, neither the BI $\rightarrow$ OOBI or TI $\rightarrow$ OOTI transitions are accompanied by gap closings. On the other hand, on the many-body level both transitions result in gap closings in PBC systems (and for the BI $\rightarrow$ OOBI in OBC systems). Subsequently, the ordering opens up the many-body gap between the two degenerate ground states and the first excited state. For large enough $V$, the single-particle gap is irrelevant, and fails to capture the many-body physics.

\section{Appendix B: Hartree Fock for Weakly Interacting Regime}

For symmetry protected topological (SPT) phases, topological classification tends to be protected from small perturbations that respect the underlying symmetry. In the SSH model considered in the main text, this discussion can be made precise by applying Hartree-Fock theory. When weak interaction (preserving chiral symmetry) is added to the model, the interaction renormalizes the inter-cell hopping strengths, i.\,e. $H(J, d, \tau, V) \approx H(J, \tilde{d}, \tilde{\tau}, V = 0)$ for new hopping parameters $\tilde{d}$, and $\tilde{\tau}$ in an effective non-interacting model. This subsection details and justifies the application of Hartree-Fock theory to the weakly-interacting model, even in finite systems with open boundary conditions.

\begin{figure}[b!]
\includegraphics[width=0.45\textwidth]{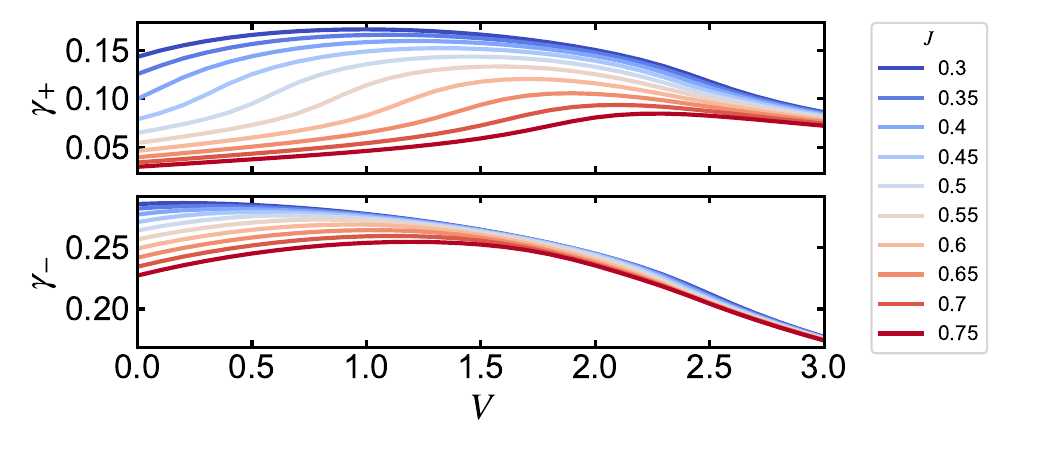}
\caption{\label{fig:gamma_pm} $\gamma_{\pm}$ which goes into calculation of $\tilde{d}$ and $\tilde{\tau}$, for system of $L = 64$ unit cells, at $d = 0.4$, $\tau = 1.0$. Weak dependence on $V$ is reflected in the approximate linearity of the BI $\rightarrow$ TI phase boundary.}
\end{figure}

The interaction term can be written explicitly in terms of sublattice creation-annihilation operators as

\begin{align}
{\hat{H}_I}& = V \sum_j \hat{n}_j^a \hat{n}_{j + 1}^b + \hat{n}_{j+1}^a \hat{n}_{j}^b \nonumber \\
& = V \sum_j \hat{a}_j^{\dag}\hat{a}_j \hat{b}_{j+1}^{\dag}\hat{b}_{j+1} + \hat{a}_{j+1}^{\dag}\hat{a}_{j+1} \hat{b}_{j}^{\dag}\hat{b}_{j},
\end{align}

In the Hartree-Fock approximation, the Hamiltonian is approximately transformed back into the non-interacting basis by decoupling the quartic interaction terms into sums of quadratic terms:

\begin{align}
{\hat{H}_{I,HF}}=  V& \sum_j \big[ \langle \hat{n}^a_j \rangle  \hat{n}_{j + 1}^b + \langle \hat{n}_{j + 1}^b\rangle \hat{n}^a_j\\
& +\, \langle \hat{n}_{j+1}^a \rangle \hat{n}_{j}^b + \langle \hat{n}_{j}^b \rangle \hat{n}_{j+1}^a \big] \nonumber \\
 - V& \sum_j \big[ \langle \hat{a}_j \hat{b}^{\dag}_{j+1}\rangle  \hat{a}^{\dag}_j \hat{b}_{j + 1} \nonumber \\
 & +\, \langle \hat{a}^{\dag}_{j+1} \hat{b}_{j}\rangle  \hat{b}^{\dag}_j \hat{a}_{j + 1} + h.c.\big] \nonumber,
\end{align}

The first summation is the Hartree term, $\hat{H}_{H}$, while the second summation is the Fock term, $\hat{H}_{F}$.

\begin{figure}[t]
\includegraphics[width=0.48\textwidth]{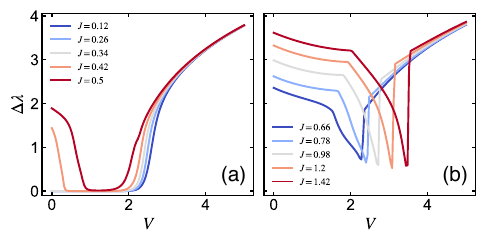}%
\caption{\label{fig:gap} Entanglement gap in symmetry broken phase as a function of interaction strength for systems of size $L = 100$ unit cells, computed with open boundary conditions, with $d = 0.4$, $\tau = 1.0$. (a) Cuts that pass through the Topological Insulating phase (b) and cuts that do not pass through topological phases. Gap opens in (a) when the chiral symmetry is spontaneously broken. In (b), phase transitions are accompanied by non-smooth changes in slope.}
\end{figure}

At half-filling, the system exhibits particle-hole symmetry. For OBC systems, AB sublattice symmetry ensures that each orbital in the bulk is near half-filling. The interaction we employ preserves particle-hole symmetry, so in the presence of weak interaction, this orbital half-filling condition should roughly hold. Thus, we expect that $\langle \hat{n}^{a,b}_j \rangle \approx \frac{1}{2}$ in the bulk for small $V$. This means that, in the thermodynamic limit, the Hartree term can be approximated to good accuracy as 
\begin{align}
{\hat{H}_{H}}  \approx  \frac{V}{2} \sum_j \big[ \hat{n}^a_j +  \hat{n}^b_j  +  \hat{n}^a_{j+1} +  \hat{n}^b_{j + 1} \big]  \approx V \hat{n}^{tot},
\end{align}

Since particle number is conserved, this quantity commutes with the Hamiltonian, and thus just adds a constant shift to the chemical potential.

Moreover, for large systems, we expect the bulk to behave in roughly translationally invariant fashion, so that terms of the form 
$\langle \hat{a}_j \hat{b}^{\dag}_{j +1}\rangle$ and $\langle \hat{a}^{\dag}_{j+1} \hat{b}_{j}\rangle$ are approximately independent of site $j$. This assumption allows us to drastically simplify the Fock term, and as a result the Hamiltonian.

Defining symmetrized and anti-symmetrized combinations of these quantities:

\begin{align}
{\gamma_{\pm}} =  \frac{\langle \hat{a}_j \hat{b}^{\dag}_{j +1}\rangle \pm \langle \hat{a}^{\dag}_{j+1} \hat{b}_{j}\rangle}{2},
\end{align}

we obtain new effective parameters, $\tilde{d}$ and $\tilde{\tau}$ such that $H(J, d, \tau, V) \approx H(J, \tilde{d}, \tilde{\tau}, V = 0)$.

\begin{figure*}[hbt!]
\includegraphics[width=\textwidth]{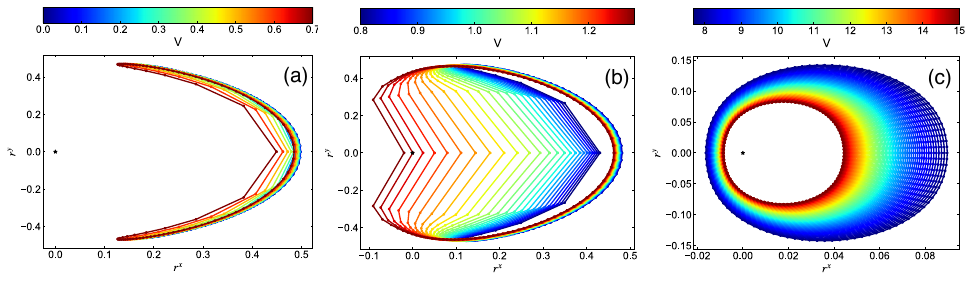}%
\caption{\label{fig:pseudospin} Pseudospin visualizations ($r^z = 0$) for systems of size $L = 100$ unit cells, computed with open boundary conditions, at $d = 0.4$, $J = 0.6$, $\tau = 1.0$ for varying interaction strength. (a) Weak interaction; Still in trivial phase (b) Intermediate interaction strength; Crossover from trivial to topological phase (c) Strong interaction; Single particle picture no longer holds, and $\langle |\hat{\mathbf{r}}| \rangle \rightarrow 0$ as in Ising ferromagnet.}
\end{figure*}

Namely, 

\begin{align}
 \tilde{d} & = d + 2 V \gamma_{+}\\
 \tilde{\tau} & =\tau + 2 V \gamma_{-}
 \label{eq:gamma_pm},
\end{align}

In Fig.~\ref{fig:gamma_pm}, $\gamma_{\pm}$ are computed from DMRG ground-state simulations for open systems according to (\ref{eq:gamma_pm}) for various values of $J$ and $V$, and we explicitly see the weak dependence of $\gamma_{+}$ on $V$ for fixed $J$. This explains the approximate linearity of the BI $\rightarrow$ TI phase boundary in the weakly interaction regime in Fig.~2~(a) in the main text and Fig.~\ref{fig:energy_diffs}. Moreover, for $V \leq 3.0$ these simulations show that the variance in $\hat{a}_j \hat{b}^{\dag}_{j +1}$ and $\hat{a}^{\dag}_{j+1} \hat{b}_{j}$ over all sites is less than $1.2\%$ of the mean value for all $J \leq 0.75$. In most cases it is substantially lower, and generically grows as $V$ increases. This means that the translation invariance assumption holds to very good approximation.

\begin{figure*}[ht!]
\includegraphics[width=\textwidth]{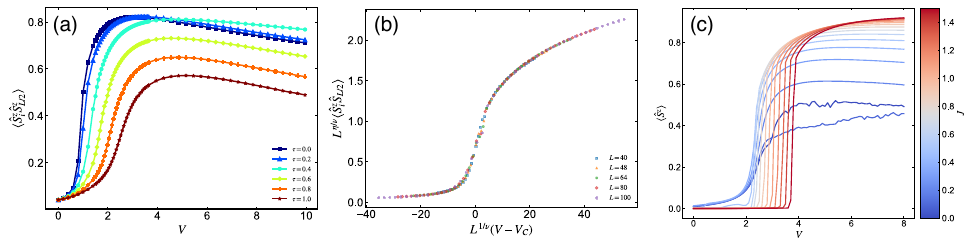}%
\caption{\label{fig:OOBI_numdens} Mean OOBI Number Density, used as an order parameter for OOBI phase transition, computed for systems with open boundary conditions. (a) $\tau$ dependence for fixed system size $L = 24$ computed at $d = 0.4, J = 0.6$ (b) Data collapse with 2d Ising critical exponents $\nu = 1, \eta = 0.25$ for systems of size $40$ to $100$ unit cells with $d = 0.4, \tau = 0, J = 0.6$ (c) Pseudospin $S^z$ in symmetry broken state.}
\end{figure*}

\section{Appendix C: Effective Spin Model for Strongly Interacting Regime}

\subsection{Case I: $d=\tau=0$}
As an initial attempt, we disregard topology, setting $d = \tau = 0$, and analyzing the low-energy physics of the resulting model. In this case, we no longer have inter-cell hopping. 
Our transformation to spin operators takes the form:

\begin{align}
\hat{S}^z_j & \leftarrow \hat{a}^{\dag}_j \hat{a}_j - \hat{b}^{\dag}_j \hat{b}_j\\
\hat{S}^{+}_j & \leftarrow \hat{a}^{\dag}_j \hat{b}_j \\
\hat{S}^{-}_j & \leftarrow \hat{b}^{\dag}_j \hat{a}_j,
\end{align}

We expand $\hat{S}^z$:

\begin{align}
\hat{S}^z_j  \hat{S}^z_{j + 1} =  (\hat{n}^a_j - \hat{n}^b_j)  (\hat{n}^a_{j+1} - \hat{n}^b_{j+1}),
\end{align}

Next, we subtract $\hat{n}^{tot}_j \hat{n}^{tot}_{j+1}$ from both sides. In the strongly interacting regime, translation symmetry in the bulk means that every unit cell has roughly one particle at half-filling. However, we demand even more strongly (a condition which is confirmed to hold true via simulation) that $\hat{n}^{tot}_j \hat{n}^{tot}_{j+1}  \approx 1$ for all $j$.

Then we can re-express the interaction as:

\begin{align}
V \sum_j [\hat{n}^b_j\hat{n}^a_{j + 1} + \hat{n}^a_j\hat{n}^b_{j + 1}] \approx  - \sum_j \frac{1}{2} \hat{S}^z_j  \hat{S}^z_{j + 1} + \mathrm{const.} \ ,
\end{align}

Under this approximation, we can rewrite the Hamiltonian explicitly in terms of spin operators:

\begin{align}
\hat{H}_{eff} = - \sum_j [ \frac{2 J}{V}\hat{S}^x_j +  \hat{S}^z_j  \hat{S}^z_{j + 1}]
\end{align}

and we see that we recover an Ising model in transverse field, with critical field strength dependent on the ratio $J/V$. This explains the approximate linearity of the OOBI $\rightarrow$ OOTI phase boundary in Fig.~2~(a) in the main text and Fig.~\ref{fig:energy_diffs}. We note that this pseudospin model holds for all values of $J$ and $V$, in the limit that $d = \tau = 0$. In addition, we see that when $V$ dominates (past the critical point), there is a two-fold degeneracy in ground state energy, corresponding to $\otimes_i \ket{\uparrow}_i$ and  $\otimes_i \ket{\downarrow}_i$, with all occupation on sublattice $\mathrm{a}$ and $\mathrm{b}$ respectively. This degeneracy is obviously not related to the topology of the non-interacting SSH model. 

Even for $d, \tau \neq 0$, we find that the system behaves like an Ising ferromagnet, with order parameter $\hat{S}^z_j = \hat{n}^a_j - \hat{n}^b_j$, as illustrated in Fig.~\ref{fig:OOBI_numdens}. In the strongly interacting phase, without broken symmetry, $\langle \hat{S}^z \rangle = 0$ (due to the double degeneracy of the ground state), but the long-range correlations $\langle \hat{S}^z_i \hat{S}^z_j  \rangle$ are non-zero. In the symmetry broken state, obtained by applying positive (negative) on-site energy terms to sublattice $\mathrm{a}$ ($\mathrm{b}$) sites, we obtain a non-zero effective magnetization. As a function of interaction strength, $V$, we find the onset of this order to be well-characterized by a second order quantum phase transition of the classical 2D Ising universality class. We demonstrate this characterization via scaling collapse with Ising critical exponents in Fig.~\ref{fig:OOBI_numdens}~(b).

\subsection{Perturbation Theory Analysis}
More formally, we can arrive at a pseudospin model by starting with ground states of the interaction term, and looking at low energy excitations that remain within the eigenspace of the interaction when perturbing $\hat{V}$ with $\hat{H}_0$.

\begin{align}
\ket{0}_a & = \prod_j  \hat{a}^{\dag}_j \ket{\emptyset} \nonumber\\
\ket{0}_b & = \prod_j  \hat{b}^{\dag}_j \ket{\emptyset},
\end{align}

where $\ket{\emptyset}$ denotes the vacuum. These are the degenerate ground states with all particles in sublattice $\mathrm{a}$ ($\mathrm{b}$). These two states have energy $E_0 = 0$ with respect to $\hat{V}$.

In the thermodynamic limit, we can ignore boundary effects, and generic first excited states take the form

\begin{align}
\ket{1}_a & = \hat{b}^{\dag}_j \hat{a}_{j\pm 1} \ket{0}_{a},
\end{align}

and similarly for $\ket{1}_b$. These states have energy $E_1 = 1$.

There are multiple ways of generating eigenstates with $E_2 = 1$. Namely, by single application of $\hat{b}^{\dag}_j \hat{a}_{j}$, or terms of the form $\hat{b}^{\dag}_j \hat{a}_{j} \hat{b}^{\dag}_k \hat{a}_{k}$. Hence, to second order, lowest-lying eigenstates of $\hat{V}$ are given by terms proportional to $J$, and to $(d \pm \tau)^2/J$.

We note that in general, other ground states of $\hat{V}$ exist, including states of the form $\prod_{j}\hat{a}^{\dag}_{2j} \hat{b}^{\dag}_{2j} \ket{\emptyset}$ (charge density waves), but the addition of $d$ and $\tau$ favors the ones considered above. We also note that while in the special case $d = \tau = 0$, there is no topology, the model can once again exhibit topology once $d$ and $\tau$ are added back in.

\section{Appendix D: Technical Details}
For DMRG simulations performed in ITensor, ground states are calculated with maximum truncation error $10^{-8}$, and noise $10^{-10}$ added to the density matrix to facilitate convergence. 

To identify the OOBI phase transition points, we compute the ground state averages for orbital charge density wave correlators for a grid of interaction strengths, for OBC systems of size $L = 48$ and $L = 64$ sites. The initial guess for critical interaction strength is computed by interpolating an intersection between these two curves. Then, an Adam optimizer is employed to minimize error with all scaled points from both system sizes fitted to one curve. The specific error function used assigns more weight to data near the critical point, reflecting the fact that the scaling should hold more closely as we approach criticality. DMRG yields very good data collapse when treated with the Ising critical exponents. We find that both qualitatively and quantitatively this holds for all $d$. In addition, for different values of $\tau$, the magnitude of the average correlations below and above the critical point change, but the phase transition behavior remains.

We note that an alternate procedure can be used to characterize the CDW phase transition when dealing with PBC systems. In this case, both the BI and TI have singly degenerate ground states, whereas both order-obstructed phases are doubly degenerate. As a result, the CDW transition can be identified by pinpointing the closing of the many-body energy gap. We have run coarse-grained simulations on PBC systems across the phase diagram, and have verified that this procedure indeed gives us qualitatively the same phase boundary.

\begin{table*}[bt!]
  \centering
  \begin{tabular}{ |p{1cm}||p{1cm}|p{1cm}|p{1cm}|p{1cm}|p{1cm}|p{1cm}|p{1cm}|}
  \hline
  \multicolumn{8}{|c|}{Pump Parameters} \\
  \hline
  Sub-figure & $J$ & $d$ & $V$ & $\Delta_0$ & $R_{\Delta}$ & $\tau_0$ & $R_{\tau}$\\
 \hline
    (a) & $2.5$ & $0.8$ & $5.0$ & $0.0$ & $0.1$ & $0.1$ & $0.2$ \\
    \hline
    (b) & $0.5$ & $0.8$ & $3.0$ & $0.0$ & $0.1$ & $0.1$ & $0.2$ \\
    \hline
    (c) & $0.2$ & $0.4$ & $0.3$ & $0.01$ & $0.0$ & $0.0$ & $0.5$ \\
    \hline
    (d) & $0.2$ & $0.4$ & $0.3$ & $0.0$ & $0.01$ & $0.0$ & $0.5$ \\
    \hline
  \end{tabular}
  \caption{\label{tab:pumps} Parameter specifications for correlation-assisted charge pumps in Fig.~\ref{fig:pumps}}
  \label{tab:1}
\end{table*}

In our investigation of charge pumping, we simulate adiabatic evolution by running independent ground state simulations for each pair of parameters $(\Delta, \tau)$ along the cycle, and then smoothly connecting this data in post-processing. The specific parameters used in the cycles in Fig.~4 in the main text are given in Table~\ref{tab:pumps}. Cycles (a)-(c) were parameterized via

\begin{align}
\Delta(\theta) & = \Delta_0 + R_{\Delta} \mathrm{sin}(\theta), \nonumber \\
\tau(\theta) & = \tau_0 + R_{\tau} \mathrm{cos}(\theta).
\end{align}

whereas cycle (d) was parameterized by

\begin{align}
\Delta(\theta) & = R_{\Delta} \mathrm{sgn}\{\mathrm{sin}(\theta)\}, \nonumber \\
\tau(\theta) & = R_{\tau} \mathrm{cos}(\theta).
\end{align}

\section{Appendix E: Analyis of the Polarization}
The Resta polarization, as defined in Eq.~(2) in the main text of our manuscript, has at its core the expectation value $\Phi = \mathrm{Tr}[\hat{\rho}\, e^{\frac{i 2 \pi}{L a_0} \hat{X}}]$. The real and imaginary parts of this quantity are both relevant, as $P = q a_0 \mathrm{arg}[\Phi]/2\pi$.

\paragraph{Weak Interaction}---
In the BI and TI phases, $\mathrm{Im}[\Phi]$ is zero up to numerical precision. $\mathrm{Re}[\Phi]$ is large in both cases; it is positive in BI, leading to a value of $P=0$, and negative in TI, leading to a value of $P=qa_0/2$. 

\begin{figure}[ht!]
\includegraphics[width=0.48\textwidth]{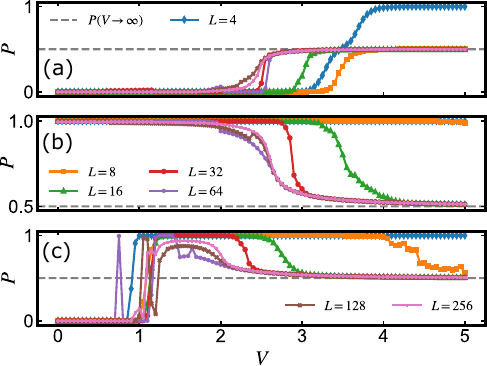}%
\caption{\label{fig:pol_scaling} Dependence of Resta polarization in symmetry broken state on system-size and interaction strength, showing that $|P| \rightarrow qa_0/4$ as $V \rightarrow \infty$, with $d = 0.4$, $\tau = 1.0$ for different values of $J$. (a) Starting from BI ($P = 0$) with $J = 1.0$, entering OOBI phase, and approaching $P=qa_0/4$ from below. (b) Starting from TI ($P = qa_0/2$) with $J = 0.1$, entering OOTI phase, and approaching $P=qa_0/4$ from above. (c) Scenario involving competition between mean-field-induced topological transition (BI $\rightarrow$ TI) and CDW transition. Starting from BI with $J = 0.6$, the state's polarization exhibits non-monotonic behavior, temporarily rising to $P=qa_0/2$ before approaching the fractionalized value $P=qa_0/4$. As can be seen in these examples, we reiterate that $\tau$ does not change the phase diagram qualitatively, although it does alter the precise locations of phase transitions.}
\end{figure}

\paragraph{Strong Interaction}---
In the $V \rightarrow \infty$ limit, $\mathrm{Re}[\Phi] = 0$ and $P = \pm qa_0/4$ identically. For finite $V$ in the CDW phase, the states have very large overlap with the $V\rightarrow \infty$ states. The two degenerate ground states in each phase have $\mathrm{Im}[\Phi]$ of equal magnitude but opposite sign, and small but non-zero $\mathrm{Re}[\Phi]$, with the sign of $\mathrm{Re}[\Phi]$ corresponding to the topology of the order-obstructed phase. In the OOBI zero-temperature ensemble (order explicitly suppressed), the large imaginary parts cancel and we recover $P = 0$. In the OOTI zero-temperature ensemble, the imaginary parts exactly cancel, but in this case the negativity of $\mathrm{Re}[\Phi]$ means that suppressing order recovers $P = qa_0/2$. 
\paragraph{Onset of Order}---
It is instructive to discuss what happens to the real and imaginary parts of $\Phi$ if we fix $J$, $d$, and $\tau$ and, starting from $V = 0$, iteratively increase $V$. Dealing with PBC systems, we know that for $V = 0$ there is one ground state, while for $V \rightarrow \infty$ there is double degeneracy. Tracing the $V = 0$ ground state as we increase $V$, $|\mathrm{Im}[\Phi]|$ increases continuously across the CDW phase transition, while $|\mathrm{Re}[\Phi]|$ shrinks. This explains how $P$ approaches $qa_0/4$ from below when starting from the BI ground state, and from above when starting with the ground state in the TI phase, as illustrated in Fig.~\ref{fig:pol_scaling}~(a) and (b) respectively. Additionally, the $\mathrm{Arg}$ in the definition of the Resta polarization means that even though the transition in $P$ is a byproduct of CDW order, $P$ scaling curves can take a vastly different shape, and will not quantitatively belong to the same universality class as the CDW order.
\paragraph{Competition with MFT Effects}---
Finally, another interesting scenario is possible according to the phase diagram. Namely, starting near the non-interacting BI $\leftrightarrow$ TI phase boundary in the trivial phase and iteratively increasing interaction, mean-field effects induce a topological transition \textit{before} the onset of order. Such a topological transition can even \textit{compete} with the CDW transition, leading to continuous yet non-monotonic behavior of the polarization as a function of $V$. Such a scenario is presented in the Fig.~\ref{fig:pol_scaling}~(c).

\section{Appendix F: Comparison of Boundary Conditions}

The Resta polarization introduced in Eq.~(2) in the main text is a many-body topological invariant defined for periodic systems. DMRG calculations, however, are far more efficient for OBC systems. It is therefore of great interest to investigate many-body topological properties in systems with open boundary conditions. In this section, we 
discuss first the OBC polarization, and second the application of the Resta polarization to OBC systems. Lastly, we discuss signatures of charge pumping in OBC systems.

\subsection{OBC Polarization}
For OBC systems, the polarization is defined as the positional expectation value of the charges, 
\begin{align}
\label{eq:pol_obc}
P_{\mathrm{OBC}}  = q\langle \hat{X} \rangle = q \sum_i x_i \langle \hat{n}_i \rangle,
\end{align}
which cannot be identified as a topological invariant. The positions $x_i$ are measured with respect to the center of the chain. Although the polarization~\eqref{eq:pol_obc} is not a strict topological invariant, a non-zero value indicates breaking of sublattice symmetry or accumulation of charge at the edges and is thus an interesting quantity to analyze.

In Fig.~\ref{fig:obc_pol} below, we have computed this OBC polarization for the ground state and first excited state (degenerate with ground state in TI and order-obstructed phases) over the entire cross-section of the phase diagram presented in Fig.~2~(a) in the main text. Although this quantity is clearly not quantized, and the values are not in one-to-one correspondence with topological character, this polarization captures some of the same qualitative behavior of the PBC Resta polarization. Generally speaking, $|\langle \hat X\rangle | \approx 1$ signifies the build-up of charge on the edges, as occurs deep in the TI phase. On the other hand, values of $|\langle \hat X\rangle | \approx 0$ are found deep within the BI phase, signifying the complete absence of charge build-up on the edges. Lastly, we see signatures of fractionalization deep in the CDW regime, where $|\langle \hat X\rangle | \approx 1/2$.

The polarization for OBC and PBC strongly differ with respect to order-obstruction. In the CDW phase, if we explicitly suppress order and 
average over the zero-temperature thermal ensemble, the PBC polarization recovers the buried topology of the underlying phase, whereas the OBC polarization gives the trivial result $P_{\mathrm{OBC}} = 0$. Intuitively, this can be understood as due to the fact that $\langle \hat{X} \rangle$ is always real-valued, while the complex exponentiation of $\hat{X}$ in the PBC polarization allows for complex phases to be combined in more ways.

\begin{figure}[ht!]
\includegraphics[width=0.48\textwidth]{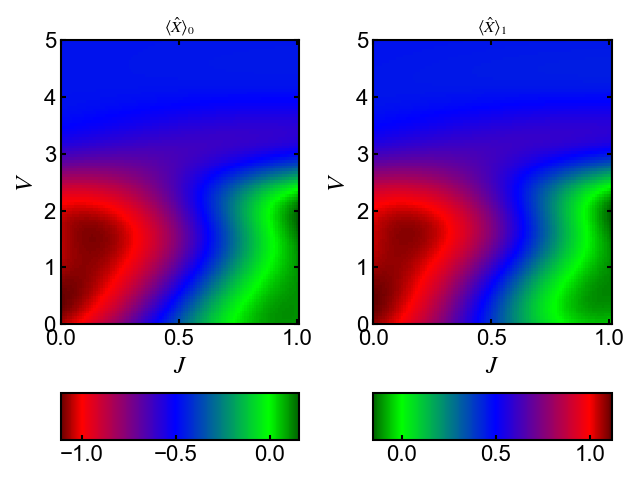}%
\caption{\label{fig:obc_pol} OBC polarization $P_{\mathrm{OBC}} = q\langle \hat{X}\rangle_{\alpha}$ with respect to ground state ($\alpha=0$) or first excited state ($\alpha=1$)
for $L = 200$ unit cells with $d = 0.4$, and $\tau = 0$. Without loss of generality, suitable choice of origin is taken so that value is system-size invariant.}
\end{figure}

\subsection{Resta Polarization in OBC Systems}
While the Resta polarization is only well-defined as a topological invariant for PBC systems, large enough OBC systems (approaching the thermodynamic limit) should exhibit the same physics, as translation-invariance is restored. One expects that the Resta polarization is appropriate for OBC systems as long as the effects of single-particle edge states are restricted to the boundaries.

\begin{figure}[b!]
\includegraphics[width=0.45\textwidth]{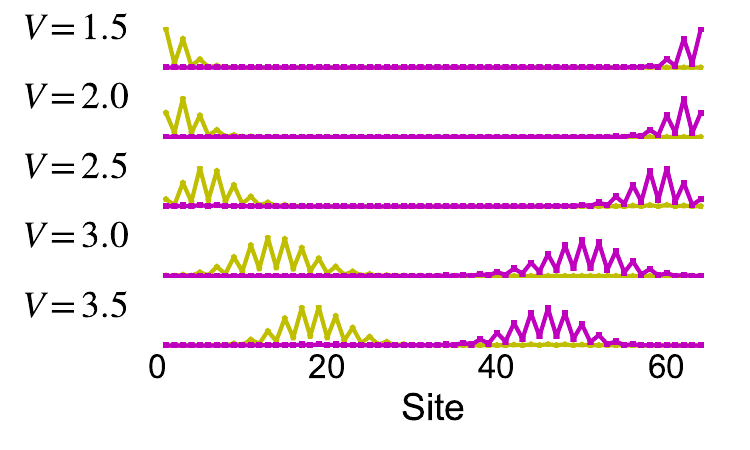}%
\caption{Natural orbital visualizations of single-particle 'edge states' with increasing $V$. $L = 64$ unit cells, at $d = 0.4$, $\tau = 1.0$.}
\label{fig:edges}
\end{figure}

Numerically, we find that in most parameter regimes considered, OBC and PBC systems yield the same value for the Resta polarization, for moderately-sized ($L \sim 100$) systems. Moreover, we verify that for $V \gg J$, (i.e. upper left corner of
Fig.~2~(a) in the main text), as $V$ increases, the single particle edge modes in OBC systems, as computed from the natural orbitals, shift away from the boundaries, as visualized in Fig.~\ref{fig:edges}. In this region (and only this region), the Resta polarization undergoes a (finite-size effect) transition in the normal phase, which should disappear in the thermodynamic limit.

Having benchmarked the similarity in behavior of OBC and PBC systems in all other regions of the phase diagram, (and having computed a course-grained phase diagram for PBC systems), we ran extensive fine-grained simulations with OBC systems, (saving roughly an order of magnitude in computational cost) the results of which are presented in Fig.~2~(a) in the main text.  

Furthermore we emphasize that, as none of the charge pump cycles for PBC systems illustrated in Fig.~4 in the main text enter the region of the phase diagram dominated by finite-size effecs, the results are qualitatively unchanged when OBC systems are used in DMRG. This means that - with obvious caveats - open-system DMRG can be used to numerically probe order-obstructed topology, potentially opening the door to huge computational savings.

\begin{figure*}[htp!]
\includegraphics[width=0.95\textwidth]{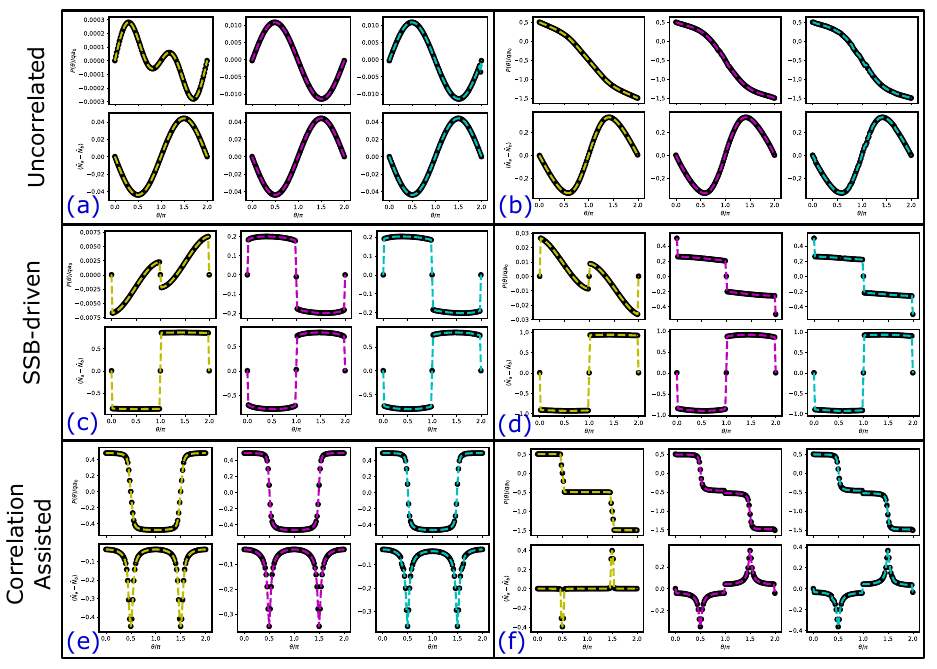} 
\caption{Full comparisons for three varieties of quantized charge pump via three different methods. Figure contains six sub-figures (a)-(f). Within each sub-figure, the left panel is obtained via MFT for periodic systems. The middle (right) panel corresponds to DMRG with open (periodic) boundary conditions. The top (bottom) row of each sub-figure plots the polarization (sublattice occupation imbalance). In the case of MFT, the polarization is computed from the Zak phase(a) and (b) depict uncorrelated pumps ($V \approx 0$) in the BI and TI phases respectively. (c)-(f) correspond to the pumps performed in Fig.~4~(a)-(d) in the main text.\label{fig:pumps_comp}}
\end{figure*}

In order to systematically corroborate our conclusions,  we have numerically simulated six varieties of charge pumps with three different approaches, with the results shown in Fig.~\ref{fig:pumps_comp}. The three approaches we consider are 1) polarization computed from the Zak phase for a periodic system in mean field theory (MFT), 2) Resta polarization for PBC systems in DMRG as presented in our manuscript, and 3) Resta polarization for the analogous OBC systems in DMRG. The six pumps we consider include the four presented in the paper, as well as one in each of the band insulating (BI) and topological insulating (TI) phases without ordering as a consistency check. 

As inferred from these additional calculations, qualitatively (also quantitatively within numerical accuracy) the DMRG results for OBC and PBC are the same. Moreover, this analysis proves quite fruitful in clarifying the distinctions between the single-particle and many-body pictures. The results of DMRG pumps for (a), (b), (e) and (f) are qualitatively consistent with MFT calculations. Quantitatively, the jumps in polarization and $\langle \hat{N}_a - \hat{N}_b\rangle$ in (e) and (f) are more abrupt in MFT, as the assumptions of MFT are closer to the thermodynamic limit. 

Pumps (c) and (d), however, are not reproducible via MFT. This is due to the fact that in the symmetry broken phase, the Zak phase (and the single-particle approximation to the polarization) approaches $\mathcal{Z} = 0$, whereas the true many-body polarization fractionalizes to $P = qa_0/4$.

\subsection{Signatures of Charge Pumping in Open Systems}

\begin{figure*}[t]
\includegraphics[width=0.85\textwidth]{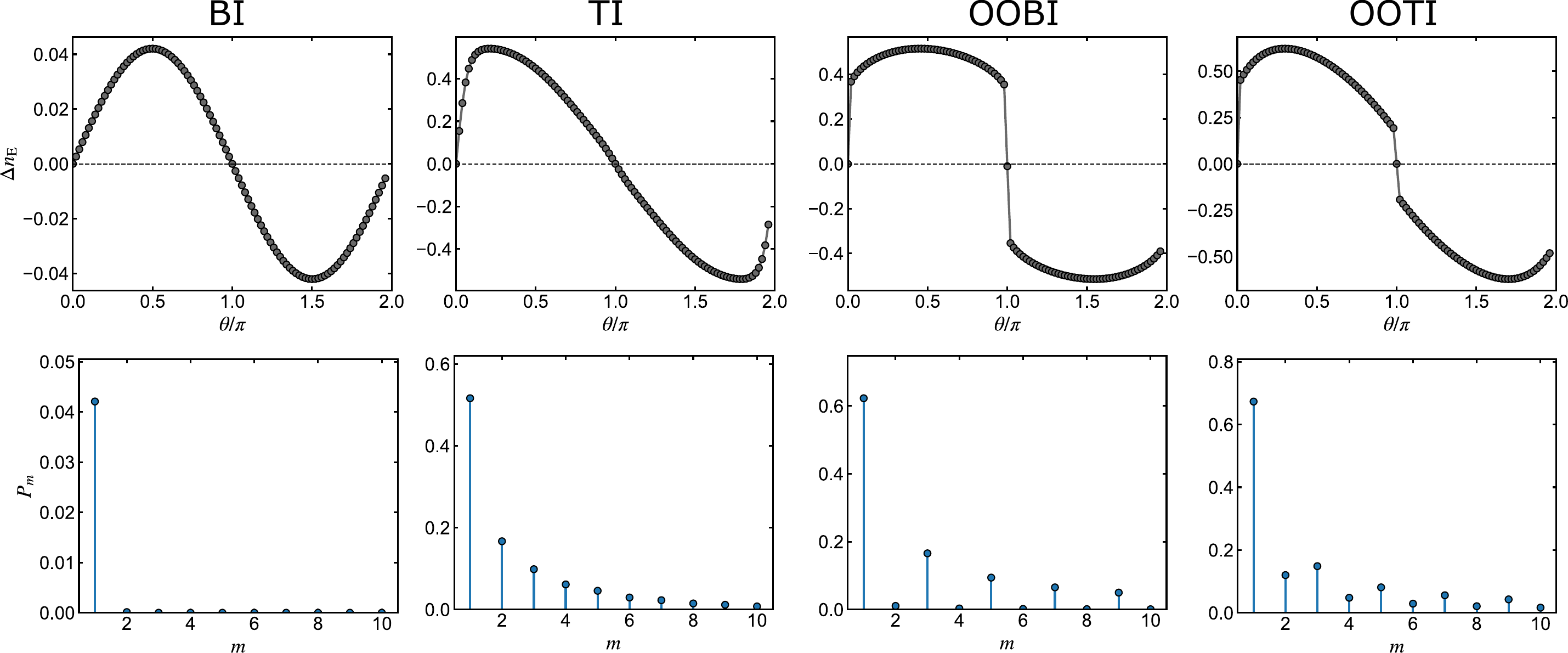}%
\caption{Top panels: Occupation difference $\Delta n_\mathrm{E}(\theta) = n_\mathrm{R}(\theta) - n_\mathrm{L}(\theta)$ for pump cycles for the uncorrelated TI, BI, OOBI (corresponding to Fig. 4 (a) in the main text), and OOTI (corresponding to Fig. 4 (b) in the main text). Bottom panels: corresponding Fourier coefficients $P_m$. \label{fig:cycle_fourier}}
\end{figure*}

Treating the system with open boundary conditions inherently asserts that the system is isolated. As such, and no charge can be pumped in OBC systems. Fu and Kane~\cite{fu_kane} argued that this constraint can be circumvented by attaching the system to a bath, which couples to the left and right edge, respectively. Adopting this argument to our system, the accumulated charge at the edges should give rise to charge transport. While performing a simulation of an open system is beyond the scope of this paper, the distinct dynamics of the charge during the pump cycle provides insights into the occurrence of charge transport in such a scenario.

\begin{figure*}[htb!]
\includegraphics[width=0.9\textwidth]{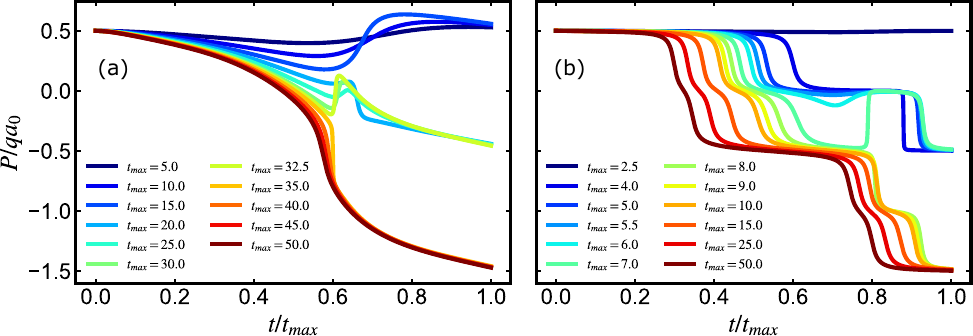}%
\caption{\label{fig:tdmrg_pumps} Explicit imaginary time-evolution for (a) uncorrelated pump (Fig.~\ref{fig:pumps_comp}~(b)) and (b) correlation-assisted pump (Fig.~\ref{fig:pumps_comp}~(f)), with $L = 8$ and periodic boundary conditions. For slow enough evolution, these cycles pump the appropriate quantized charge. Harnessing correlations, as in (b), allows one to expedite the pumping process.}
\end{figure*}

To this end, we have computed the occupation at the boundaries, $n_\mathrm{L} = \langle\hat{n}^a_1 \rangle$ and $n_\mathrm{R} = \langle\hat{n}^b_L \rangle$ from the DMRG calculations under OBC. The difference $\Delta n_\mathrm{E} = n_L - n_R$, which is an indicator for charge accumulation on the edges, is shown in Fig.~\ref{fig:cycle_fourier}. Let us discuss the TI case first. Although $\Delta n_\mathrm{E}(\theta=0) = \Delta n_\mathrm{E}(\theta=2\pi) = 0$, there is clear tendency to decrease $\Delta n_\mathrm{E}(\theta)$, i.\, e. to pump charge from right to left. For $\frac{\pi}{4} < \theta < \frac{7\pi}{4}$ the $\Delta n_\mathrm{E}(\theta)$ is qualitatively very similar to the polarization in the PBC calculation (see Fig.~\ref{fig:pumps_comp}(b)). The reflection of this charge drift at the boundaries in the OBC would be absent in an open system, thus giving rise to charge transport. Also note that $\theta$-dependence $\Delta n_\mathrm{E}(\theta)$ matches the polarization in Fig.~\ref{fig:pumps_comp}(a).

Apart from the visual tendency to move charge, we can formulate an indicative criterion that captures thus asymmetry. In particular, we notice $\Delta n_\mathrm{E}(\theta+\pi) = -\Delta n_\mathrm{E}(\theta)$ for the cycle performed in the BI phase. This property corresponds to a reflection symmetry around the $\theta=\pi$ (the middle point of the cycle) while replacing $a \leftrightarrow b$ sublattice sites. This symmetry is broken if net charge is pumped, as also inferred from Fig.~\ref{fig:pumps_comp}. Breaking this symmetry can be quantified by calculating the Fourier components
\begin{align}
    \label{eq:fourier_coeff}
    P_m = \int^{2\pi}_0 d\theta\, \sin(m\theta) \Delta n_\mathrm{E}(\theta) \ .
\end{align}
$\Delta n_\mathrm{E}(\theta+\pi) = -\Delta n_\mathrm{E}(\theta)$ leads to $P_m=0$ for even $m$. We have calculated the Fourier coefficients~\eqref{eq:fourier_coeff} for the discussed pump cycles (bottom panels in Fig.~\ref{fig:cycle_fourier}). As expected, $P_{2m} \ne 0$ indicates topological behavior, where charge would be transported in an open-system setup. Note that this criterion also holds for the polarization with PBC.

\section{Appendix G: Explicit Time Evolution}

In the main text, and in the rest of the Supplementary Material, all of the Thouless pump cycles presented are generated by performing independent ground state calculations for instantaneous Hamiltonians along the cycle, and smoothly stitching them together in post-processing. Explicitly performing these cycles in tDMRG (time-dependent DMRG) with large PBC systems is computationally intractable. However, we have performed the imaginary time evolution for such cycles with small systems of $L = 8$ unit cells. These simulations were performed in DMRG using a time-dependent Hamiltonian under second-order discretization and a maximum allowed bond dimension $m_{max} = 1000$. The time step was fixed at $\delta t = 0.005$ for all cycles, with the cycle being divided into $N$ steps. The number of steps $N$ directly parameterizes the total time $t_{max}$ taken by the cycle. By sweeping over various values of $N$ (resp. $t_{max}$), we can determine how slowly the cycle must be performed to successfully transport charge. The results of these cycles for the correlation-assisted pump in Fig.~4~(d) in the main text (as well as Fig.~\ref{fig:pumps_comp}~(f)), and the uncorrelated reference pump in Fig.~\ref{fig:pumps_comp}~(b) above, are presented in Fig.~\ref{fig:tdmrg_pumps}. Two observations can be readily made after looking at these results. First, they are in complete agreement with the results presented in the main text. And second, the correlation-assisted pump can be performed almost an order of magnitude more quickly than the uncorrelated pump, relaxing but not eliminating the adiabatic constraint.


\begin{thebibliography}{69}%
\makeatletter
\providecommand \@ifxundefined [1]{%
 \@ifx{#1\undefined}
}%
\providecommand \@ifnum [1]{%
 \ifnum #1\expandafter \@firstoftwo
 \else \expandafter \@secondoftwo
 \fi
}%
\providecommand \@ifx [1]{%
 \ifx #1\expandafter \@firstoftwo
 \else \expandafter \@secondoftwo
 \fi
}%
\providecommand \natexlab [1]{#1}%
\providecommand \enquote  [1]{``#1''}%
\providecommand \bibnamefont  [1]{#1}%
\providecommand \bibfnamefont [1]{#1}%
\providecommand \citenamefont [1]{#1}%
\providecommand \href@noop [0]{\@secondoftwo}%
\providecommand \href [0]{\begingroup \@sanitize@url \@href}%
\providecommand \@href[1]{\@@startlink{#1}\@@href}%
\providecommand \@@href[1]{\endgroup#1\@@endlink}%
\providecommand \@sanitize@url [0]{\catcode `\\12\catcode `\$12\catcode
  `\&12\catcode `\#12\catcode `\^12\catcode `\_12\catcode `\%12\relax}%
\providecommand \@@startlink[1]{}%
\providecommand \@@endlink[0]{}%
\providecommand \url  [0]{\begingroup\@sanitize@url \@url }%
\providecommand \@url [1]{\endgroup\@href {#1}{\urlprefix }}%
\providecommand \urlprefix  [0]{URL }%
\providecommand \Eprint [0]{\href }%
\providecommand \doibase [0]{https://doi.org/}%
\providecommand \selectlanguage [0]{\@gobble}%
\providecommand \bibinfo  [0]{\@secondoftwo}%
\providecommand \bibfield  [0]{\@secondoftwo}%
\providecommand \translation [1]{[#1]}%
\providecommand \BibitemOpen [0]{}%
\providecommand \bibitemStop [0]{}%
\providecommand \bibitemNoStop [0]{.\EOS\space}%
\providecommand \EOS [0]{\spacefactor3000\relax}%
\providecommand \BibitemShut  [1]{\csname bibitem#1\endcsname}%
\let\auto@bib@innerbib\@empty
\bibitem [{\citenamefont {Hasan}\ and\ \citenamefont
  {Kane}(2010)}]{hasan_colloquium:_2010}%
  \BibitemOpen
  \bibfield  {author} {\bibinfo {author} {\bibfnamefont {M.~Z.}\ \bibnamefont
  {Hasan}}\ and\ \bibinfo {author} {\bibfnamefont {C.~L.}\ \bibnamefont
  {Kane}},\ }\href {https://doi.org/10.1103/RevModPhys.82.3045} {\bibfield
  {journal} {\bibinfo  {journal} {Rev. Mod. Phys.}\ }\textbf {\bibinfo {volume}
  {82}},\ \bibinfo {pages} {3045} (\bibinfo {year} {2010})}\BibitemShut
  {NoStop}%
\bibitem [{\citenamefont {Qi}\ and\ \citenamefont {Zhang}(2011)}]{qi_rmp_2011}%
  \BibitemOpen
  \bibfield  {author} {\bibinfo {author} {\bibfnamefont {X.-L.}\ \bibnamefont
  {Qi}}\ and\ \bibinfo {author} {\bibfnamefont {S.-C.}\ \bibnamefont {Zhang}},\
  }\href {https://doi.org/10.1103/RevModPhys.83.1057} {\bibfield  {journal}
  {\bibinfo  {journal} {Rev. Mod. Phys.}\ }\textbf {\bibinfo {volume} {83}},\
  \bibinfo {pages} {1057} (\bibinfo {year} {2011})}\BibitemShut {NoStop}%
\bibitem [{\citenamefont {Chang}\ \emph {et~al.}(2013)\citenamefont {Chang},
  \citenamefont {Zhang}, \citenamefont {Feng}, \citenamefont {Shen},
  \citenamefont {Zhang}, \citenamefont {Guo}, \citenamefont {Li}, \citenamefont
  {Ou}, \citenamefont {Wei}, \citenamefont {Wang}, \citenamefont {Ji},
  \citenamefont {Feng}, \citenamefont {Ji}, \citenamefont {Chen}, \citenamefont
  {Jia}, \citenamefont {Dai}, \citenamefont {Fang}, \citenamefont {Zhang},
  \citenamefont {He}, \citenamefont {Wang}, \citenamefont {Lu}, \citenamefont
  {Ma},\ and\ \citenamefont {Xue}}]{chang_experimental_2013}%
  \BibitemOpen
  \bibfield  {author} {\bibinfo {author} {\bibfnamefont {C.-Z.}\ \bibnamefont
  {Chang}}, \bibinfo {author} {\bibfnamefont {J.}~\bibnamefont {Zhang}},
  \bibinfo {author} {\bibfnamefont {X.}~\bibnamefont {Feng}}, \bibinfo {author}
  {\bibfnamefont {J.}~\bibnamefont {Shen}}, \bibinfo {author} {\bibfnamefont
  {Z.}~\bibnamefont {Zhang}}, \bibinfo {author} {\bibfnamefont
  {M.}~\bibnamefont {Guo}}, \bibinfo {author} {\bibfnamefont {K.}~\bibnamefont
  {Li}}, \bibinfo {author} {\bibfnamefont {Y.}~\bibnamefont {Ou}}, \bibinfo
  {author} {\bibfnamefont {P.}~\bibnamefont {Wei}}, \bibinfo {author}
  {\bibfnamefont {L.-L.}\ \bibnamefont {Wang}}, \bibinfo {author}
  {\bibfnamefont {Z.-Q.}\ \bibnamefont {Ji}}, \bibinfo {author} {\bibfnamefont
  {Y.}~\bibnamefont {Feng}}, \bibinfo {author} {\bibfnamefont {S.}~\bibnamefont
  {Ji}}, \bibinfo {author} {\bibfnamefont {X.}~\bibnamefont {Chen}}, \bibinfo
  {author} {\bibfnamefont {J.}~\bibnamefont {Jia}}, \bibinfo {author}
  {\bibfnamefont {X.}~\bibnamefont {Dai}}, \bibinfo {author} {\bibfnamefont
  {Z.}~\bibnamefont {Fang}}, \bibinfo {author} {\bibfnamefont {S.-C.}\
  \bibnamefont {Zhang}}, \bibinfo {author} {\bibfnamefont {K.}~\bibnamefont
  {He}}, \bibinfo {author} {\bibfnamefont {Y.}~\bibnamefont {Wang}}, \bibinfo
  {author} {\bibfnamefont {L.}~\bibnamefont {Lu}}, \bibinfo {author}
  {\bibfnamefont {X.-C.}\ \bibnamefont {Ma}},\ and\ \bibinfo {author}
  {\bibfnamefont {Q.-K.}\ \bibnamefont {Xue}},\ }\href
  {https://doi.org/10.1126/science.1234414} {\bibfield  {journal} {\bibinfo
  {journal} {Science}\ }\textbf {\bibinfo {volume} {340}},\ \bibinfo {pages}
  {167} (\bibinfo {year} {2013})}\BibitemShut {NoStop}%
\bibitem [{\citenamefont {K\"{o}nig}\ \emph {et~al.}(2007)\citenamefont
  {K\"{o}nig}, \citenamefont {Wiedmann}, \citenamefont {Br\"{u}ne},
  \citenamefont {Roth}, \citenamefont {Buhmann}, \citenamefont {Molenkamp},
  \citenamefont {Qi},\ and\ \citenamefont {Zhang}}]{konig_quantum_2007}%
  \BibitemOpen
  \bibfield  {author} {\bibinfo {author} {\bibfnamefont {M.}~\bibnamefont
  {K\"{o}nig}}, \bibinfo {author} {\bibfnamefont {S.}~\bibnamefont {Wiedmann}},
  \bibinfo {author} {\bibfnamefont {C.}~\bibnamefont {Br\"{u}ne}}, \bibinfo
  {author} {\bibfnamefont {A.}~\bibnamefont {Roth}}, \bibinfo {author}
  {\bibfnamefont {H.}~\bibnamefont {Buhmann}}, \bibinfo {author} {\bibfnamefont
  {L.~W.}\ \bibnamefont {Molenkamp}}, \bibinfo {author} {\bibfnamefont {X.-L.}\
  \bibnamefont {Qi}},\ and\ \bibinfo {author} {\bibfnamefont {S.-C.}\
  \bibnamefont {Zhang}},\ }\href {https://doi.org/10.1126/science.1148047}
  {\bibfield  {journal} {\bibinfo  {journal} {Science}\ }\textbf {\bibinfo
  {volume} {318}},\ \bibinfo {pages} {766} (\bibinfo {year}
  {2007})}\BibitemShut {NoStop}%
\bibitem [{\citenamefont {Yue}\ \emph {et~al.}(2017)\citenamefont {Yue},
  \citenamefont {Xue}, \citenamefont {Liu}, \citenamefont {Wang},\ and\
  \citenamefont {Gu}}]{yue_nanometric_2017}%
  \BibitemOpen
  \bibfield  {author} {\bibinfo {author} {\bibfnamefont {Z.}~\bibnamefont
  {Yue}}, \bibinfo {author} {\bibfnamefont {G.}~\bibnamefont {Xue}}, \bibinfo
  {author} {\bibfnamefont {J.}~\bibnamefont {Liu}}, \bibinfo {author}
  {\bibfnamefont {Y.}~\bibnamefont {Wang}},\ and\ \bibinfo {author}
  {\bibfnamefont {M.}~\bibnamefont {Gu}},\ }\href
  {https://doi.org/10.1038/ncomms15354} {\bibfield  {journal} {\bibinfo
  {journal} {Nat Commun}\ }\textbf {\bibinfo {volume} {8}},\ \bibinfo {pages}
  {1} (\bibinfo {year} {2017})}\BibitemShut {NoStop}%
\bibitem [{\citenamefont {Klitzing}\ \emph {et~al.}(1980)\citenamefont
  {Klitzing}, \citenamefont {Dorda},\ and\ \citenamefont
  {Pepper}}]{klitzing_prl_1980}%
  \BibitemOpen
  \bibfield  {author} {\bibinfo {author} {\bibfnamefont {K.~v.}\ \bibnamefont
  {Klitzing}}, \bibinfo {author} {\bibfnamefont {G.}~\bibnamefont {Dorda}},\
  and\ \bibinfo {author} {\bibfnamefont {M.}~\bibnamefont {Pepper}},\ }\href
  {https://doi.org/10.1103/PhysRevLett.45.494} {\bibfield  {journal} {\bibinfo
  {journal} {Phys. Rev. Lett.}\ }\textbf {\bibinfo {volume} {45}},\ \bibinfo
  {pages} {494} (\bibinfo {year} {1980})}\BibitemShut {NoStop}%
\bibitem [{\citenamefont {Prange}\ and\ \citenamefont
  {Girvin}(1987)}]{prange_qhe_1987}%
  \BibitemOpen
  \bibfield  {author} {\bibinfo {author} {\bibfnamefont {R.}~\bibnamefont
  {Prange}}\ and\ \bibinfo {author} {\bibfnamefont {S.}~\bibnamefont
  {Girvin}},\ }\href@noop {} {\emph {\bibinfo {title} {The Quantum Hall
  effect}}},\ Graduate texts in contemporary physics\ (\bibinfo  {publisher}
  {Springer-Verlag},\ \bibinfo {year} {1987})\BibitemShut {NoStop}%
\bibitem [{\citenamefont {Thouless}\ \emph {et~al.}(1982)\citenamefont
  {Thouless}, \citenamefont {Kohmoto}, \citenamefont {Nightingale},\ and\
  \citenamefont {den Nijs}}]{tknn_1982}%
  \BibitemOpen
  \bibfield  {author} {\bibinfo {author} {\bibfnamefont {D.~J.}\ \bibnamefont
  {Thouless}}, \bibinfo {author} {\bibfnamefont {M.}~\bibnamefont {Kohmoto}},
  \bibinfo {author} {\bibfnamefont {M.~P.}\ \bibnamefont {Nightingale}},\ and\
  \bibinfo {author} {\bibfnamefont {M.}~\bibnamefont {den Nijs}},\ }\href
  {https://doi.org/10.1103/PhysRevLett.49.405} {\bibfield  {journal} {\bibinfo
  {journal} {Phys. Rev. Lett.}\ }\textbf {\bibinfo {volume} {49}},\ \bibinfo
  {pages} {405} (\bibinfo {year} {1982})}\BibitemShut {NoStop}%
\bibitem [{\citenamefont {Laughlin}(1981)}]{laughlin_1981}%
  \BibitemOpen
  \bibfield  {author} {\bibinfo {author} {\bibfnamefont {R.~B.}\ \bibnamefont
  {Laughlin}},\ }\href {https://doi.org/10.1103/PhysRevB.23.5632} {\bibfield
  {journal} {\bibinfo  {journal} {Phys. Rev. B}\ }\textbf {\bibinfo {volume}
  {23}},\ \bibinfo {pages} {5632} (\bibinfo {year} {1981})}\BibitemShut
  {NoStop}%
\bibitem [{\citenamefont {Thouless}(1983)}]{thouless_prb_1983}%
  \BibitemOpen
  \bibfield  {author} {\bibinfo {author} {\bibfnamefont {D.~J.}\ \bibnamefont
  {Thouless}},\ }\href {https://doi.org/10.1103/PhysRevB.27.6083} {\bibfield
  {journal} {\bibinfo  {journal} {Phys. Rev. B}\ }\textbf {\bibinfo {volume}
  {27}},\ \bibinfo {pages} {6083} (\bibinfo {year} {1983})}\BibitemShut
  {NoStop}%
\bibitem [{\citenamefont {Haldane}(1988)}]{haldane_model_1988}%
  \BibitemOpen
  \bibfield  {author} {\bibinfo {author} {\bibfnamefont {F.~D.~M.}\
  \bibnamefont {Haldane}},\ }\href
  {https://doi.org/10.1103/PhysRevLett.61.2015} {\bibfield  {journal} {\bibinfo
   {journal} {Phys. Rev. Lett.}\ }\textbf {\bibinfo {volume} {61}},\ \bibinfo
  {pages} {2015} (\bibinfo {year} {1988})}\BibitemShut {NoStop}%
\bibitem [{\citenamefont {Qi}\ \emph {et~al.}(2008)\citenamefont {Qi},
  \citenamefont {Hughes},\ and\ \citenamefont {Zhang}}]{qi_topological_2008}%
  \BibitemOpen
  \bibfield  {author} {\bibinfo {author} {\bibfnamefont {X.-L.}\ \bibnamefont
  {Qi}}, \bibinfo {author} {\bibfnamefont {T.~L.}\ \bibnamefont {Hughes}},\
  and\ \bibinfo {author} {\bibfnamefont {S.-C.}\ \bibnamefont {Zhang}},\ }\href
  {https://doi.org/10.1103/PhysRevB.78.195424} {\bibfield  {journal} {\bibinfo
  {journal} {Phys. Rev. B}\ }\textbf {\bibinfo {volume} {78}},\ \bibinfo
  {pages} {195424} (\bibinfo {year} {2008})}\BibitemShut {NoStop}%
\bibitem [{\citenamefont {Rachel}\ \emph {et~al.}(2010)\citenamefont {Rachel},
  \citenamefont {Schuricht}, \citenamefont {Scharfenberger}, \citenamefont
  {Thomale},\ and\ \citenamefont {Greiter}}]{Rachel_2010}%
  \BibitemOpen
  \bibfield  {author} {\bibinfo {author} {\bibfnamefont {S.}~\bibnamefont
  {Rachel}}, \bibinfo {author} {\bibfnamefont {D.}~\bibnamefont {Schuricht}},
  \bibinfo {author} {\bibfnamefont {B.}~\bibnamefont {Scharfenberger}},
  \bibinfo {author} {\bibfnamefont {R.}~\bibnamefont {Thomale}},\ and\ \bibinfo
  {author} {\bibfnamefont {M.}~\bibnamefont {Greiter}},\ }\href
  {https://doi.org/10.1088/1742-6596/200/2/022049} {\bibfield  {journal}
  {\bibinfo  {journal} {Journal of Physics: Conference Series}\ }\textbf
  {\bibinfo {volume} {200}},\ \bibinfo {pages} {022049} (\bibinfo {year}
  {2010})}\BibitemShut {NoStop}%
\bibitem [{\citenamefont {Hohenadler}\ \emph {et~al.}(2011)\citenamefont
  {Hohenadler}, \citenamefont {Lang},\ and\ \citenamefont
  {Assaad}}]{Hohenadler_prl_2011}%
  \BibitemOpen
  \bibfield  {author} {\bibinfo {author} {\bibfnamefont {M.}~\bibnamefont
  {Hohenadler}}, \bibinfo {author} {\bibfnamefont {T.~C.}\ \bibnamefont
  {Lang}},\ and\ \bibinfo {author} {\bibfnamefont {F.~F.}\ \bibnamefont
  {Assaad}},\ }\href {https://doi.org/10.1103/PhysRevLett.106.100403}
  {\bibfield  {journal} {\bibinfo  {journal} {Phys. Rev. Lett.}\ }\textbf
  {\bibinfo {volume} {106}},\ \bibinfo {pages} {100403} (\bibinfo {year}
  {2011})}\BibitemShut {NoStop}%
\bibitem [{\citenamefont {Budich}\ \emph {et~al.}(2012)\citenamefont {Budich},
  \citenamefont {Thomale}, \citenamefont {Li}, \citenamefont {Laubach},\ and\
  \citenamefont {Zhang}}]{budich_fluctuation-induced_2012}%
  \BibitemOpen
  \bibfield  {author} {\bibinfo {author} {\bibfnamefont {J.~C.}\ \bibnamefont
  {Budich}}, \bibinfo {author} {\bibfnamefont {R.}~\bibnamefont {Thomale}},
  \bibinfo {author} {\bibfnamefont {G.}~\bibnamefont {Li}}, \bibinfo {author}
  {\bibfnamefont {M.}~\bibnamefont {Laubach}},\ and\ \bibinfo {author}
  {\bibfnamefont {S.-C.}\ \bibnamefont {Zhang}},\ }\href
  {https://doi.org/10.1103/PhysRevB.86.201407} {\bibfield  {journal} {\bibinfo
  {journal} {Phys. Rev. B}\ }\textbf {\bibinfo {volume} {86}},\ \bibinfo
  {pages} {201407(R)} (\bibinfo {year} {2012})}\BibitemShut {NoStop}%
\bibitem [{\citenamefont {Witczak-Krempa}\ \emph {et~al.}(2014)\citenamefont
  {Witczak-Krempa}, \citenamefont {Knap},\ and\ \citenamefont
  {Abanin}}]{kka_prl_2014}%
  \BibitemOpen
  \bibfield  {author} {\bibinfo {author} {\bibfnamefont {W.}~\bibnamefont
  {Witczak-Krempa}}, \bibinfo {author} {\bibfnamefont {M.}~\bibnamefont
  {Knap}},\ and\ \bibinfo {author} {\bibfnamefont {D.}~\bibnamefont {Abanin}},\
  }\href {https://doi.org/10.1103/PhysRevLett.113.136402} {\bibfield  {journal}
  {\bibinfo  {journal} {Phys. Rev. Lett.}\ }\textbf {\bibinfo {volume} {113}},\
  \bibinfo {pages} {136402} (\bibinfo {year} {2014})}\BibitemShut {NoStop}%
\bibitem [{\citenamefont {Amaricci}\ \emph {et~al.}(2015)\citenamefont
  {Amaricci}, \citenamefont {Budich}, \citenamefont {Capone}, \citenamefont
  {Trauzettel},\ and\ \citenamefont {Sangiovanni}}]{Amaricci_prl_2015}%
  \BibitemOpen
  \bibfield  {author} {\bibinfo {author} {\bibfnamefont {A.}~\bibnamefont
  {Amaricci}}, \bibinfo {author} {\bibfnamefont {J.~C.}\ \bibnamefont
  {Budich}}, \bibinfo {author} {\bibfnamefont {M.}~\bibnamefont {Capone}},
  \bibinfo {author} {\bibfnamefont {B.}~\bibnamefont {Trauzettel}},\ and\
  \bibinfo {author} {\bibfnamefont {G.}~\bibnamefont {Sangiovanni}},\ }\href
  {https://doi.org/10.1103/PhysRevLett.114.185701} {\bibfield  {journal}
  {\bibinfo  {journal} {Phys. Rev. Lett.}\ }\textbf {\bibinfo {volume} {114}},\
  \bibinfo {pages} {185701} (\bibinfo {year} {2015})}\BibitemShut {NoStop}%
\bibitem [{\citenamefont {Wen}(2017)}]{wen_rmp_2017}%
  \BibitemOpen
  \bibfield  {author} {\bibinfo {author} {\bibfnamefont {X.-G.}\ \bibnamefont
  {Wen}},\ }\href {https://doi.org/10.1103/RevModPhys.89.041004} {\bibfield
  {journal} {\bibinfo  {journal} {Rev. Mod. Phys.}\ }\textbf {\bibinfo {volume}
  {89}},\ \bibinfo {pages} {041004} (\bibinfo {year} {2017})}\BibitemShut
  {NoStop}%
\bibitem [{\citenamefont {Resta}(1998)}]{resta_quantum-mechanical_1998}%
  \BibitemOpen
  \bibfield  {author} {\bibinfo {author} {\bibfnamefont {R.}~\bibnamefont
  {Resta}},\ }\href {https://doi.org/10.1103/PhysRevLett.80.1800} {\bibfield
  {journal} {\bibinfo  {journal} {Phys. Rev. Lett.}\ }\textbf {\bibinfo
  {volume} {80}},\ \bibinfo {pages} {1800} (\bibinfo {year}
  {1998})}\BibitemShut {NoStop}%
\bibitem [{\citenamefont {Niu}\ and\ \citenamefont
  {Thouless}(1984)}]{Niu_1984}%
  \BibitemOpen
  \bibfield  {author} {\bibinfo {author} {\bibfnamefont {Q.}~\bibnamefont
  {Niu}}\ and\ \bibinfo {author} {\bibfnamefont {D.~J.}\ \bibnamefont
  {Thouless}},\ }\href {https://doi.org/10.1088/0305-4470/17/12/016} {\bibfield
   {journal} {\bibinfo  {journal} {J. Phys. A: Math. Gen.}\ }\textbf {\bibinfo
  {volume} {17}},\ \bibinfo {pages} {2453} (\bibinfo {year}
  {1984})}\BibitemShut {NoStop}%
\bibitem [{\citenamefont {Chen}\ \emph {et~al.}(2013)\citenamefont {Chen},
  \citenamefont {Gu}, \citenamefont {Liu},\ and\ \citenamefont
  {Wen}}]{chen_spt_prb_2013}%
  \BibitemOpen
  \bibfield  {author} {\bibinfo {author} {\bibfnamefont {X.}~\bibnamefont
  {Chen}}, \bibinfo {author} {\bibfnamefont {Z.-C.}\ \bibnamefont {Gu}},
  \bibinfo {author} {\bibfnamefont {Z.-X.}\ \bibnamefont {Liu}},\ and\ \bibinfo
  {author} {\bibfnamefont {X.-G.}\ \bibnamefont {Wen}},\ }\href
  {https://doi.org/10.1103/PhysRevB.87.155114} {\bibfield  {journal} {\bibinfo
  {journal} {Phys. Rev. B}\ }\textbf {\bibinfo {volume} {87}},\ \bibinfo
  {pages} {155114} (\bibinfo {year} {2013})}\BibitemShut {NoStop}%
\bibitem [{\citenamefont {Gonz\'alez-Cuadra}\ \emph {et~al.}(2019)\citenamefont
  {Gonz\'alez-Cuadra}, \citenamefont {Dauphin}, \citenamefont {Grzybowski},
  \citenamefont {W\'ojcik}, \citenamefont {Lewenstein},\ and\ \citenamefont
  {Bermudez}}]{cuadra_prb_2019}%
  \BibitemOpen
  \bibfield  {author} {\bibinfo {author} {\bibfnamefont {D.}~\bibnamefont
  {Gonz\'alez-Cuadra}}, \bibinfo {author} {\bibfnamefont {A.}~\bibnamefont
  {Dauphin}}, \bibinfo {author} {\bibfnamefont {P.~R.}\ \bibnamefont
  {Grzybowski}}, \bibinfo {author} {\bibfnamefont {P.}~\bibnamefont
  {W\'ojcik}}, \bibinfo {author} {\bibfnamefont {M.}~\bibnamefont
  {Lewenstein}},\ and\ \bibinfo {author} {\bibfnamefont {A.}~\bibnamefont
  {Bermudez}},\ }\href {https://doi.org/10.1103/PhysRevB.99.045139} {\bibfield
  {journal} {\bibinfo  {journal} {Phys. Rev. B}\ }\textbf {\bibinfo {volume}
  {99}},\ \bibinfo {pages} {045139} (\bibinfo {year} {2019})}\BibitemShut
  {NoStop}%
\bibitem [{\citenamefont {Gonz\'{a}lez-Cuadra}\ \emph
  {et~al.}(2019{\natexlab{a}})\citenamefont {Gonz\'{a}lez-Cuadra},
  \citenamefont {Bermudez}, \citenamefont {Grzybowski}, \citenamefont
  {Lewenstein},\ and\ \citenamefont {Dauphin}}]{cuadra_natcomm_2019}%
  \BibitemOpen
  \bibfield  {author} {\bibinfo {author} {\bibfnamefont {D.}~\bibnamefont
  {Gonz\'{a}lez-Cuadra}}, \bibinfo {author} {\bibfnamefont {A.}~\bibnamefont
  {Bermudez}}, \bibinfo {author} {\bibfnamefont {P.~R.}\ \bibnamefont
  {Grzybowski}}, \bibinfo {author} {\bibfnamefont {M.}~\bibnamefont
  {Lewenstein}},\ and\ \bibinfo {author} {\bibfnamefont {A.}~\bibnamefont
  {Dauphin}},\ }\bibfield  {journal} {\bibinfo  {journal} {Nature
  Communications}\ }\textbf {\bibinfo {volume} {10}},\ \href
  {https://doi.org/10.1038/s41467-019-10796-8} {10.1038/s41467-019-10796-8}
  (\bibinfo {year} {2019}{\natexlab{a}})\BibitemShut {NoStop}%
\bibitem [{\citenamefont {Gonz\'{a}lez-Cuadra}\ \emph
  {et~al.}(2019{\natexlab{b}})\citenamefont {Gonz\'{a}lez-Cuadra},
  \citenamefont {Dauphin}, \citenamefont {Grzybowski}, \citenamefont
  {Lewenstein},\ and\ \citenamefont {Bermudez}}]{cuadra_soliton_2019}%
  \BibitemOpen
  \bibfield  {author} {\bibinfo {author} {\bibfnamefont {D.}~\bibnamefont
  {Gonz\'{a}lez-Cuadra}}, \bibinfo {author} {\bibfnamefont {A.}~\bibnamefont
  {Dauphin}}, \bibinfo {author} {\bibfnamefont {P.~R.}\ \bibnamefont
  {Grzybowski}}, \bibinfo {author} {\bibfnamefont {M.}~\bibnamefont
  {Lewenstein}},\ and\ \bibinfo {author} {\bibfnamefont {A.}~\bibnamefont
  {Bermudez}},\ }\href@noop {} {\bibinfo {title} {$\mathbb{Z}_n$ solitons in
  intertwined topological phases}} (\bibinfo {year} {2019}{\natexlab{b}}),\
  \Eprint {https://arxiv.org/abs/1908.02186} {arXiv:1908.02186
  [cond-mat.quant-gas]} \BibitemShut {NoStop}%
\bibitem [{\citenamefont {Marra}\ \emph {et~al.}(2015)\citenamefont {Marra},
  \citenamefont {Citro},\ and\ \citenamefont {Ortix}}]{PhysRevB.91.125411}%
  \BibitemOpen
  \bibfield  {author} {\bibinfo {author} {\bibfnamefont {P.}~\bibnamefont
  {Marra}}, \bibinfo {author} {\bibfnamefont {R.}~\bibnamefont {Citro}},\ and\
  \bibinfo {author} {\bibfnamefont {C.}~\bibnamefont {Ortix}},\ }\href
  {https://doi.org/10.1103/PhysRevB.91.125411} {\bibfield  {journal} {\bibinfo
  {journal} {Phys. Rev. B}\ }\textbf {\bibinfo {volume} {91}},\ \bibinfo
  {pages} {125411} (\bibinfo {year} {2015})}\BibitemShut {NoStop}%
\bibitem [{\citenamefont {Li}\ and\ \citenamefont
  {Fleischhauer}(2017)}]{PhysRevB.96.085444}%
  \BibitemOpen
  \bibfield  {author} {\bibinfo {author} {\bibfnamefont {R.}~\bibnamefont
  {Li}}\ and\ \bibinfo {author} {\bibfnamefont {M.}~\bibnamefont
  {Fleischhauer}},\ }\href {https://doi.org/10.1103/PhysRevB.96.085444}
  {\bibfield  {journal} {\bibinfo  {journal} {Phys. Rev. B}\ }\textbf {\bibinfo
  {volume} {96}},\ \bibinfo {pages} {085444} (\bibinfo {year}
  {2017})}\BibitemShut {NoStop}%
\bibitem [{\citenamefont {Het\'enyi}(2020)}]{Hetenyi_2020}%
  \BibitemOpen
  \bibfield  {author} {\bibinfo {author} {\bibfnamefont {B.}~\bibnamefont
  {Het\'enyi}},\ }\bibfield  {journal} {\bibinfo  {journal} {Physical Review
  Research}\ }\textbf {\bibinfo {volume} {2}},\ \href
  {https://doi.org/10.1103/physrevresearch.2.023277}
  {10.1103/physrevresearch.2.023277} (\bibinfo {year} {2020})\BibitemShut
  {NoStop}%
\bibitem [{\citenamefont {White}(1992)}]{white_dmrg_1992}%
  \BibitemOpen
  \bibfield  {author} {\bibinfo {author} {\bibfnamefont {S.~R.}\ \bibnamefont
  {White}},\ }\href {https://doi.org/10.1103/PhysRevLett.69.2863} {\bibfield
  {journal} {\bibinfo  {journal} {Phys. Rev. Lett.}\ }\textbf {\bibinfo
  {volume} {69}},\ \bibinfo {pages} {2863} (\bibinfo {year}
  {1992})}\BibitemShut {NoStop}%
\bibitem [{\citenamefont
  {Schollw\"{o}ck}(2005)}]{schollwock_density-matrix_2005}%
  \BibitemOpen
  \bibfield  {author} {\bibinfo {author} {\bibfnamefont {U.}~\bibnamefont
  {Schollw\"{o}ck}},\ }\href {https://doi.org/10.1103/RevModPhys.77.259}
  {\bibfield  {journal} {\bibinfo  {journal} {Rev. Mod. Phys.}\ }\textbf
  {\bibinfo {volume} {77}},\ \bibinfo {pages} {259} (\bibinfo {year}
  {2005})}\BibitemShut {NoStop}%
\bibitem [{\citenamefont {Su}\ \emph {et~al.}(1979)\citenamefont {Su},
  \citenamefont {Schrieffer},\ and\ \citenamefont {Heeger}}]{su_solitons_1979}%
  \BibitemOpen
  \bibfield  {author} {\bibinfo {author} {\bibfnamefont {W.~P.}\ \bibnamefont
  {Su}}, \bibinfo {author} {\bibfnamefont {J.~R.}\ \bibnamefont {Schrieffer}},\
  and\ \bibinfo {author} {\bibfnamefont {A.~J.}\ \bibnamefont {Heeger}},\
  }\href {https://doi.org/10.1103/PhysRevLett.42.1698} {\bibfield  {journal}
  {\bibinfo  {journal} {Phys. Rev. Lett.}\ }\textbf {\bibinfo {volume} {42}},\
  \bibinfo {pages} {1698} (\bibinfo {year} {1979})}\BibitemShut {NoStop}%
\bibitem [{\citenamefont {Heeger}\ \emph {et~al.}(1988)\citenamefont {Heeger},
  \citenamefont {Kivelson}, \citenamefont {Schrieffer},\ and\ \citenamefont
  {Su}}]{heeger_solitons_1988}%
  \BibitemOpen
  \bibfield  {author} {\bibinfo {author} {\bibfnamefont {A.~J.}\ \bibnamefont
  {Heeger}}, \bibinfo {author} {\bibfnamefont {S.}~\bibnamefont {Kivelson}},
  \bibinfo {author} {\bibfnamefont {J.~R.}\ \bibnamefont {Schrieffer}},\ and\
  \bibinfo {author} {\bibfnamefont {W.~P.}\ \bibnamefont {Su}},\ }\href
  {https://doi.org/10.1103/RevModPhys.60.781} {\bibfield  {journal} {\bibinfo
  {journal} {Rev. Mod. Phys.}\ }\textbf {\bibinfo {volume} {60}},\ \bibinfo
  {pages} {781} (\bibinfo {year} {1988})}\BibitemShut {NoStop}%
\bibitem [{\citenamefont {Aidelsburger}\ \emph {et~al.}(2011)\citenamefont
  {Aidelsburger}, \citenamefont {Atala}, \citenamefont {Nascimb\`{e}ne},
  \citenamefont {Trotzky}, \citenamefont {Chen},\ and\ \citenamefont
  {Bloch}}]{aidelsburger_experimental_2011}%
  \BibitemOpen
  \bibfield  {author} {\bibinfo {author} {\bibfnamefont {M.}~\bibnamefont
  {Aidelsburger}}, \bibinfo {author} {\bibfnamefont {M.}~\bibnamefont {Atala}},
  \bibinfo {author} {\bibfnamefont {S.}~\bibnamefont {Nascimb\`{e}ne}},
  \bibinfo {author} {\bibfnamefont {S.}~\bibnamefont {Trotzky}}, \bibinfo
  {author} {\bibfnamefont {Y.-A.}\ \bibnamefont {Chen}},\ and\ \bibinfo
  {author} {\bibfnamefont {I.}~\bibnamefont {Bloch}},\ }\href
  {https://doi.org/10.1103/PhysRevLett.107.255301} {\bibfield  {journal}
  {\bibinfo  {journal} {Phys. Rev. Lett.}\ }\textbf {\bibinfo {volume} {107}},\
  \bibinfo {pages} {255301} (\bibinfo {year} {2011})}\BibitemShut {NoStop}%
\bibitem [{\citenamefont {Struck}\ \emph {et~al.}(2012)\citenamefont {Struck},
  \citenamefont {\"{O}lschl\"{a}ger}, \citenamefont {Weinberg}, \citenamefont
  {Hauke}, \citenamefont {Simonet}, \citenamefont {Eckardt}, \citenamefont
  {Lewenstein}, \citenamefont {Sengstock},\ and\ \citenamefont
  {Windpassinger}}]{struck_tunable_2012}%
  \BibitemOpen
  \bibfield  {author} {\bibinfo {author} {\bibfnamefont {J.}~\bibnamefont
  {Struck}}, \bibinfo {author} {\bibfnamefont {C.}~\bibnamefont
  {\"{O}lschl\"{a}ger}}, \bibinfo {author} {\bibfnamefont {M.}~\bibnamefont
  {Weinberg}}, \bibinfo {author} {\bibfnamefont {P.}~\bibnamefont {Hauke}},
  \bibinfo {author} {\bibfnamefont {J.}~\bibnamefont {Simonet}}, \bibinfo
  {author} {\bibfnamefont {A.}~\bibnamefont {Eckardt}}, \bibinfo {author}
  {\bibfnamefont {M.}~\bibnamefont {Lewenstein}}, \bibinfo {author}
  {\bibfnamefont {K.}~\bibnamefont {Sengstock}},\ and\ \bibinfo {author}
  {\bibfnamefont {P.}~\bibnamefont {Windpassinger}},\ }\href
  {https://doi.org/10.1103/PhysRevLett.108.225304} {\bibfield  {journal}
  {\bibinfo  {journal} {Phys. Rev. Lett.}\ }\textbf {\bibinfo {volume} {108}},\
  \bibinfo {pages} {225304} (\bibinfo {year} {2012})}\BibitemShut {NoStop}%
\bibitem [{\citenamefont {Goldman}\ \emph {et~al.}(2016)\citenamefont
  {Goldman}, \citenamefont {Budich},\ and\ \citenamefont
  {Zoller}}]{goldman_topological_2016}%
  \BibitemOpen
  \bibfield  {author} {\bibinfo {author} {\bibfnamefont {N.}~\bibnamefont
  {Goldman}}, \bibinfo {author} {\bibfnamefont {J.~C.}\ \bibnamefont
  {Budich}},\ and\ \bibinfo {author} {\bibfnamefont {P.}~\bibnamefont
  {Zoller}},\ }\href {https://doi.org/10.1038/nphys3803} {\bibfield  {journal}
  {\bibinfo  {journal} {Nature Phys.}\ }\textbf {\bibinfo {volume} {12}},\
  \bibinfo {pages} {639} (\bibinfo {year} {2016})}\BibitemShut {NoStop}%
\bibitem [{\citenamefont {Dauphin}\ \emph {et~al.}(2017)\citenamefont
  {Dauphin}, \citenamefont {Tran}, \citenamefont {Lewenstein},\ and\
  \citenamefont {Goldman}}]{dauphin_loading_2017}%
  \BibitemOpen
  \bibfield  {author} {\bibinfo {author} {\bibfnamefont {A.}~\bibnamefont
  {Dauphin}}, \bibinfo {author} {\bibfnamefont {D.-T.}\ \bibnamefont {Tran}},
  \bibinfo {author} {\bibfnamefont {M.}~\bibnamefont {Lewenstein}},\ and\
  \bibinfo {author} {\bibfnamefont {N.}~\bibnamefont {Goldman}},\ }\href
  {https://doi.org/10.1088/2053-1583/aa6a3b} {\bibfield  {journal} {\bibinfo
  {journal} {2D Mater.}\ }\textbf {\bibinfo {volume} {4}},\ \bibinfo {pages}
  {024010} (\bibinfo {year} {2017})}\BibitemShut {NoStop}%
\bibitem [{\citenamefont {Fl\"{a}schner}\ \emph {et~al.}(2016)\citenamefont
  {Fl\"{a}schner}, \citenamefont {Rem}, \citenamefont {Tarnowski},
  \citenamefont {Vogel}, \citenamefont {L\"{u}hmann}, \citenamefont
  {Sengstock},\ and\ \citenamefont {Weitenberg}}]{flaschner_experimental_2016}%
  \BibitemOpen
  \bibfield  {author} {\bibinfo {author} {\bibfnamefont {N.}~\bibnamefont
  {Fl\"{a}schner}}, \bibinfo {author} {\bibfnamefont {B.~S.}\ \bibnamefont
  {Rem}}, \bibinfo {author} {\bibfnamefont {M.}~\bibnamefont {Tarnowski}},
  \bibinfo {author} {\bibfnamefont {D.}~\bibnamefont {Vogel}}, \bibinfo
  {author} {\bibfnamefont {D.-S.}\ \bibnamefont {L\"{u}hmann}}, \bibinfo
  {author} {\bibfnamefont {K.}~\bibnamefont {Sengstock}},\ and\ \bibinfo
  {author} {\bibfnamefont {C.}~\bibnamefont {Weitenberg}},\ }\href
  {https://doi.org/10.1126/science.aad4568} {\bibfield  {journal} {\bibinfo
  {journal} {Science}\ }\textbf {\bibinfo {volume} {352}},\ \bibinfo {pages}
  {1091} (\bibinfo {year} {2016})}\BibitemShut {NoStop}%
\bibitem [{\citenamefont {Lohse}\ \emph {et~al.}(2016)\citenamefont {Lohse},
  \citenamefont {Schweizer}, \citenamefont {Zilberberg}, \citenamefont
  {Aidelsburger},\ and\ \citenamefont {Bloch}}]{lohse_thouless_2016}%
  \BibitemOpen
  \bibfield  {author} {\bibinfo {author} {\bibfnamefont {M.}~\bibnamefont
  {Lohse}}, \bibinfo {author} {\bibfnamefont {C.}~\bibnamefont {Schweizer}},
  \bibinfo {author} {\bibfnamefont {O.}~\bibnamefont {Zilberberg}}, \bibinfo
  {author} {\bibfnamefont {M.}~\bibnamefont {Aidelsburger}},\ and\ \bibinfo
  {author} {\bibfnamefont {I.}~\bibnamefont {Bloch}},\ }\href
  {https://doi.org/10.1038/nphys3584} {\bibfield  {journal} {\bibinfo
  {journal} {Nature Phys.}\ }\textbf {\bibinfo {volume} {12}},\ \bibinfo
  {pages} {350} (\bibinfo {year} {2016})}\BibitemShut {NoStop}%
\bibitem [{\citenamefont {Nakajima}\ \emph {et~al.}(2016)\citenamefont
  {Nakajima}, \citenamefont {Tomita}, \citenamefont {Taie}, \citenamefont
  {Ichinose}, \citenamefont {Ozawa}, \citenamefont {Wang}, \citenamefont
  {Troyer},\ and\ \citenamefont {Takahashi}}]{nakajima_topological_2016}%
  \BibitemOpen
  \bibfield  {author} {\bibinfo {author} {\bibfnamefont {S.}~\bibnamefont
  {Nakajima}}, \bibinfo {author} {\bibfnamefont {T.}~\bibnamefont {Tomita}},
  \bibinfo {author} {\bibfnamefont {S.}~\bibnamefont {Taie}}, \bibinfo {author}
  {\bibfnamefont {T.}~\bibnamefont {Ichinose}}, \bibinfo {author}
  {\bibfnamefont {H.}~\bibnamefont {Ozawa}}, \bibinfo {author} {\bibfnamefont
  {L.}~\bibnamefont {Wang}}, \bibinfo {author} {\bibfnamefont {M.}~\bibnamefont
  {Troyer}},\ and\ \bibinfo {author} {\bibfnamefont {Y.}~\bibnamefont
  {Takahashi}},\ }\href {https://doi.org/10.1038/nphys3622} {\bibfield
  {journal} {\bibinfo  {journal} {Nature Phys.}\ }\textbf {\bibinfo {volume}
  {12}},\ \bibinfo {pages} {296} (\bibinfo {year} {2016})}\BibitemShut
  {NoStop}%
\bibitem [{\citenamefont {J\"{u}nemann}\ \emph {et~al.}(2017)\citenamefont
  {J\"{u}nemann}, \citenamefont {Piga}, \citenamefont {Ran}, \citenamefont
  {Lewenstein}, \citenamefont {Rizzi},\ and\ \citenamefont
  {Bermudez}}]{junemann_exploring_2017}%
  \BibitemOpen
  \bibfield  {author} {\bibinfo {author} {\bibfnamefont {J.}~\bibnamefont
  {J\"{u}nemann}}, \bibinfo {author} {\bibfnamefont {A.}~\bibnamefont {Piga}},
  \bibinfo {author} {\bibfnamefont {S.-J.}\ \bibnamefont {Ran}}, \bibinfo
  {author} {\bibfnamefont {M.}~\bibnamefont {Lewenstein}}, \bibinfo {author}
  {\bibfnamefont {M.}~\bibnamefont {Rizzi}},\ and\ \bibinfo {author}
  {\bibfnamefont {A.}~\bibnamefont {Bermudez}},\ }\href
  {https://doi.org/10.1103/PhysRevX.7.031057} {\bibfield  {journal} {\bibinfo
  {journal} {Phys. Rev. X}\ }\textbf {\bibinfo {volume} {7}},\ \bibinfo {pages}
  {031057} (\bibinfo {year} {2017})}\BibitemShut {NoStop}%
\bibitem [{\citenamefont {Sbierski}\ and\ \citenamefont
  {Karrasch}(2018)}]{sbierski_topological_2018}%
  \BibitemOpen
  \bibfield  {author} {\bibinfo {author} {\bibfnamefont {B.}~\bibnamefont
  {Sbierski}}\ and\ \bibinfo {author} {\bibfnamefont {C.}~\bibnamefont
  {Karrasch}},\ }\href {https://doi.org/10.1103/PhysRevB.98.165101} {\bibfield
  {journal} {\bibinfo  {journal} {Phys. Rev. B}\ }\textbf {\bibinfo {volume}
  {98}},\ \bibinfo {pages} {165101} (\bibinfo {year} {2018})}\BibitemShut
  {NoStop}%
\bibitem [{\citenamefont {Marques}\ and\ \citenamefont
  {Dias}(2017)}]{marques_multihole_2017}%
  \BibitemOpen
  \bibfield  {author} {\bibinfo {author} {\bibfnamefont {A.~M.}\ \bibnamefont
  {Marques}}\ and\ \bibinfo {author} {\bibfnamefont {R.~G.}\ \bibnamefont
  {Dias}},\ }\href {https://doi.org/10.1103/PhysRevB.95.115443} {\bibfield
  {journal} {\bibinfo  {journal} {Phys. Rev. B}\ }\textbf {\bibinfo {volume}
  {95}},\ \bibinfo {pages} {115443} (\bibinfo {year} {2017})}\BibitemShut
  {NoStop}%
\bibitem [{\citenamefont {Anisimovas}\ \emph {et~al.}(2016)\citenamefont
  {Anisimovas}, \citenamefont {Ra\v{c}i\={u}nas}, \citenamefont {Str\"{a}ter},
  \citenamefont {Eckardt}, \citenamefont {Spielman},\ and\ \citenamefont
  {Juzeli\={u}nas}}]{anisimovas_semisynthetic_2016}%
  \BibitemOpen
  \bibfield  {author} {\bibinfo {author} {\bibfnamefont {E.}~\bibnamefont
  {Anisimovas}}, \bibinfo {author} {\bibfnamefont {M.}~\bibnamefont
  {Ra\v{c}i\={u}nas}}, \bibinfo {author} {\bibfnamefont {C.}~\bibnamefont
  {Str\"{a}ter}}, \bibinfo {author} {\bibfnamefont {A.}~\bibnamefont
  {Eckardt}}, \bibinfo {author} {\bibfnamefont {I.~B.}\ \bibnamefont
  {Spielman}},\ and\ \bibinfo {author} {\bibfnamefont {G.}~\bibnamefont
  {Juzeli\={u}nas}},\ }\href {https://doi.org/10.1103/PhysRevA.94.063632}
  {\bibfield  {journal} {\bibinfo  {journal} {Phys. Rev. A}\ }\textbf {\bibinfo
  {volume} {94}},\ \bibinfo {pages} {063632} (\bibinfo {year}
  {2016})}\BibitemShut {NoStop}%
\bibitem [{\citenamefont {Kruckenhauser}\ and\ \citenamefont
  {Budich}(2018)}]{kruckenhauser_dynamical_2018}%
  \BibitemOpen
  \bibfield  {author} {\bibinfo {author} {\bibfnamefont {A.}~\bibnamefont
  {Kruckenhauser}}\ and\ \bibinfo {author} {\bibfnamefont {J.~C.}\ \bibnamefont
  {Budich}},\ }\href {https://doi.org/10.1103/PhysRevB.98.195124} {\bibfield
  {journal} {\bibinfo  {journal} {Phys. Rev. B}\ }\textbf {\bibinfo {volume}
  {98}},\ \bibinfo {pages} {195124} (\bibinfo {year} {2018})}\BibitemShut
  {NoStop}%
\bibitem [{the()}]{thermalizing}%
  \BibitemOpen
  \bibinfo {note} {Unlike the intracell interaction $\hat{U} = U \sum_j
  \hat{n}^a_j \hat{n}^b_j$, $\hat{V}$ is efficient at thermalizing, which is
  important for dynamically preparing such phases.}\BibitemShut {Stop}%
\bibitem [{\citenamefont {Yahyavi}\ \emph {et~al.}(2018)\citenamefont
  {Yahyavi}, \citenamefont {Saleem},\ and\ \citenamefont
  {Hetényi}}]{Yahyavi_2018}%
  \BibitemOpen
  \bibfield  {author} {\bibinfo {author} {\bibfnamefont {M.}~\bibnamefont
  {Yahyavi}}, \bibinfo {author} {\bibfnamefont {L.}~\bibnamefont {Saleem}},\
  and\ \bibinfo {author} {\bibfnamefont {B.}~\bibnamefont {Hetényi}},\ }\href
  {https://doi.org/10.1088/1361-648x/aae0a4} {\bibfield  {journal} {\bibinfo
  {journal} {Journal of Physics: Condensed Matter}\ }\textbf {\bibinfo {volume}
  {30}},\ \bibinfo {pages} {445602} (\bibinfo {year} {2018})}\BibitemShut
  {NoStop}%
\bibitem [{Note1()}]{Note1}%
  \BibitemOpen
  \bibinfo {note} {Our analysis strictly excludes the point $J = 0$, at which
  the model obeys different symmetries.}\BibitemShut {Stop}%
\bibitem [{sup()}]{supplement}%
  \BibitemOpen
  \bibinfo {note} {Supplemental material, available online.}\BibitemShut
  {Stop}%
\bibitem [{\citenamefont {Bardyn}\ \emph {et~al.}(2018)\citenamefont {Bardyn},
  \citenamefont {Wawer}, \citenamefont {Altland}, \citenamefont
  {Fleischhauer},\ and\ \citenamefont {Diehl}}]{bardyn_probing_2018}%
  \BibitemOpen
  \bibfield  {author} {\bibinfo {author} {\bibfnamefont {C.-E.}\ \bibnamefont
  {Bardyn}}, \bibinfo {author} {\bibfnamefont {L.}~\bibnamefont {Wawer}},
  \bibinfo {author} {\bibfnamefont {A.}~\bibnamefont {Altland}}, \bibinfo
  {author} {\bibfnamefont {M.}~\bibnamefont {Fleischhauer}},\ and\ \bibinfo
  {author} {\bibfnamefont {S.}~\bibnamefont {Diehl}},\ }\href
  {https://doi.org/10.1103/PhysRevX.8.011035} {\bibfield  {journal} {\bibinfo
  {journal} {Phys. Rev. X}\ }\textbf {\bibinfo {volume} {8}},\ \bibinfo {pages}
  {011035} (\bibinfo {year} {2018})}\BibitemShut {NoStop}%
\bibitem [{\citenamefont {Fu}\ and\ \citenamefont {Kane}(2006)}]{fu_kane}%
  \BibitemOpen
  \bibfield  {author} {\bibinfo {author} {\bibfnamefont {L.}~\bibnamefont
  {Fu}}\ and\ \bibinfo {author} {\bibfnamefont {C.~L.}\ \bibnamefont {Kane}},\
  }\href {https://doi.org/10.1103/PhysRevB.74.195312} {\bibfield  {journal}
  {\bibinfo  {journal} {Phys. Rev. B}\ }\textbf {\bibinfo {volume} {74}},\
  \bibinfo {pages} {195312} (\bibinfo {year} {2006})}\BibitemShut {NoStop}%
\bibitem [{\citenamefont {Altland}\ and\ \citenamefont
  {Zirnbauer}(1997)}]{altland_nonstandard_1997}%
  \BibitemOpen
  \bibfield  {author} {\bibinfo {author} {\bibfnamefont {A.}~\bibnamefont
  {Altland}}\ and\ \bibinfo {author} {\bibfnamefont {M.~R.}\ \bibnamefont
  {Zirnbauer}},\ }\href {https://doi.org/10.1103/PhysRevB.55.1142} {\bibfield
  {journal} {\bibinfo  {journal} {Phys. Rev. B}\ }\textbf {\bibinfo {volume}
  {55}},\ \bibinfo {pages} {1142} (\bibinfo {year} {1997})}\BibitemShut
  {NoStop}%
\bibitem [{Note2()}]{Note2}%
  \BibitemOpen
  \bibinfo {note} {$P$ defined in this way is only strictly valid as a
  topological invariant for periodic systems. The polarization for OBC systems
  is defined as $P_{OBC} = q\langle \protect \hat {X}\rangle $. However, $P$ is
  still well-defined for OBC systems as long as the effects of single particle
  edge modes are restricted to the boundaries.}\BibitemShut {Stop}%
\bibitem [{\citenamefont {Pollmann}\ \emph {et~al.}(2010)\citenamefont
  {Pollmann}, \citenamefont {Turner}, \citenamefont {Berg},\ and\ \citenamefont
  {Oshikawa}}]{pollmann_entanglement_2010}%
  \BibitemOpen
  \bibfield  {author} {\bibinfo {author} {\bibfnamefont {F.}~\bibnamefont
  {Pollmann}}, \bibinfo {author} {\bibfnamefont {A.~M.}\ \bibnamefont
  {Turner}}, \bibinfo {author} {\bibfnamefont {E.}~\bibnamefont {Berg}},\ and\
  \bibinfo {author} {\bibfnamefont {M.}~\bibnamefont {Oshikawa}},\ }\href
  {https://doi.org/10.1103/PhysRevB.81.064439} {\bibfield  {journal} {\bibinfo
  {journal} {Phys. Rev. B}\ }\textbf {\bibinfo {volume} {81}},\ \bibinfo
  {pages} {064439} (\bibinfo {year} {2010})}\BibitemShut {NoStop}%
\bibitem [{\citenamefont {Fidkowski}\ and\ \citenamefont
  {Kitaev}(2011)}]{fidkowski_topological_2011}%
  \BibitemOpen
  \bibfield  {author} {\bibinfo {author} {\bibfnamefont {L.}~\bibnamefont
  {Fidkowski}}\ and\ \bibinfo {author} {\bibfnamefont {A.}~\bibnamefont
  {Kitaev}},\ }\href {https://doi.org/10.1103/PhysRevB.83.075103} {\bibfield
  {journal} {\bibinfo  {journal} {Phys. Rev. B}\ }\textbf {\bibinfo {volume}
  {83}},\ \bibinfo {pages} {075103} (\bibinfo {year} {2011})}\BibitemShut
  {NoStop}%
\bibitem [{\citenamefont {Li}\ and\ \citenamefont
  {Haldane}(2008)}]{li_entanglement_2008}%
  \BibitemOpen
  \bibfield  {author} {\bibinfo {author} {\bibfnamefont {H.}~\bibnamefont
  {Li}}\ and\ \bibinfo {author} {\bibfnamefont {F.~D.~M.}\ \bibnamefont
  {Haldane}},\ }\href {https://doi.org/10.1103/PhysRevLett.101.010504}
  {\bibfield  {journal} {\bibinfo  {journal} {Phys. Rev. Lett.}\ }\textbf
  {\bibinfo {volume} {101}},\ \bibinfo {pages} {010504} (\bibinfo {year}
  {2008})}\BibitemShut {NoStop}%
\bibitem [{\citenamefont {Zaletel}\ \emph {et~al.}(2014)\citenamefont
  {Zaletel}, \citenamefont {Mong},\ and\ \citenamefont
  {Pollmann}}]{Zaletel_2014}%
  \BibitemOpen
  \bibfield  {author} {\bibinfo {author} {\bibfnamefont {M.~P.}\ \bibnamefont
  {Zaletel}}, \bibinfo {author} {\bibfnamefont {R.~S.~K.}\ \bibnamefont
  {Mong}},\ and\ \bibinfo {author} {\bibfnamefont {F.}~\bibnamefont
  {Pollmann}},\ }\href {https://doi.org/10.1088/1742-5468/2014/10/p10007}
  {\bibfield  {journal} {\bibinfo  {journal} {Journal of Statistical Mechanics:
  Theory and Experiment}\ }\textbf {\bibinfo {volume} {2014}},\ \bibinfo
  {pages} {P10007} (\bibinfo {year} {2014})}\BibitemShut {NoStop}%
\bibitem [{\citenamefont {Hayward}\ \emph {et~al.}(2018)\citenamefont
  {Hayward}, \citenamefont {Schweizer}, \citenamefont {Lohse}, \citenamefont
  {Aidelsburger},\ and\ \citenamefont
  {Heidrich-Meisner}}]{hayward_topological_2018}%
  \BibitemOpen
  \bibfield  {author} {\bibinfo {author} {\bibfnamefont {A.}~\bibnamefont
  {Hayward}}, \bibinfo {author} {\bibfnamefont {C.}~\bibnamefont {Schweizer}},
  \bibinfo {author} {\bibfnamefont {M.}~\bibnamefont {Lohse}}, \bibinfo
  {author} {\bibfnamefont {M.}~\bibnamefont {Aidelsburger}},\ and\ \bibinfo
  {author} {\bibfnamefont {F.}~\bibnamefont {Heidrich-Meisner}},\ }\href
  {https://doi.org/10.1103/PhysRevB.98.245148} {\bibfield  {journal} {\bibinfo
  {journal} {Phys. Rev. B}\ }\textbf {\bibinfo {volume} {98}},\ \bibinfo
  {pages} {245148} (\bibinfo {year} {2018})}\BibitemShut {NoStop}%
\bibitem [{\citenamefont {Ryu}\ and\ \citenamefont
  {Hatsugai}(2006)}]{Ryu_Hatsugai_2006}%
  \BibitemOpen
  \bibfield  {author} {\bibinfo {author} {\bibfnamefont {S.}~\bibnamefont
  {Ryu}}\ and\ \bibinfo {author} {\bibfnamefont {Y.}~\bibnamefont {Hatsugai}},\
  }\href {https://doi.org/10.1103/PhysRevB.73.245115} {\bibfield  {journal}
  {\bibinfo  {journal} {Phys. Rev. B}\ }\textbf {\bibinfo {volume} {73}},\
  \bibinfo {pages} {245115} (\bibinfo {year} {2006})}\BibitemShut {NoStop}%
\bibitem [{Note3()}]{Note3}%
  \BibitemOpen
  \bibinfo {note} {This is true for PBC systems, where only the OOTI and OOBI
  phases have doubly degenerate ground states. For OBC systems the TI phase
  does as well, as illustrated by the blue line in \ref
  {fig:phase_diagram}~(a). More sophisticated measures must be taken for OBC
  systems, as detailed in the Supplementary Material~\cite
  {supplement}}\BibitemShut {NoStop}%
\bibitem [{\citenamefont {P\'{e}rez-Gonz\'{a}lez}\ \emph
  {et~al.}(2018)\citenamefont {P\'{e}rez-Gonz\'{a}lez}, \citenamefont {Bello},
  \citenamefont {G\'{o}mez-Le\'{o}n},\ and\ \citenamefont
  {Platero}}]{perez-gonzalez_ssh_2018}%
  \BibitemOpen
  \bibfield  {author} {\bibinfo {author} {\bibfnamefont {B.}~\bibnamefont
  {P\'{e}rez-Gonz\'{a}lez}}, \bibinfo {author} {\bibfnamefont {M.}~\bibnamefont
  {Bello}}, \bibinfo {author} {\bibfnamefont {A.}~\bibnamefont
  {G\'{o}mez-Le\'{o}n}},\ and\ \bibinfo {author} {\bibfnamefont
  {G.}~\bibnamefont {Platero}},\ }\href {http://arxiv.org/abs/1802.03973}
  {\bibfield  {journal} {\bibinfo  {journal} {arXiv:1802.03973 [cond-mat]}\ }
  (\bibinfo {year} {2018})}\BibitemShut {NoStop}%
\bibitem [{Note4()}]{Note4}%
  \BibitemOpen
  \bibinfo {note} {In equilibrium, $\protect \mathcal {H}_k$ is proportional to
  the single-particle Hamiltonian.}\BibitemShut {Stop}%
\bibitem [{\citenamefont {Zak}(1989)}]{zak_berrys_1989}%
  \BibitemOpen
  \bibfield  {author} {\bibinfo {author} {\bibfnamefont {J.}~\bibnamefont
  {Zak}},\ }\href {https://doi.org/10.1103/PhysRevLett.62.2747} {\bibfield
  {journal} {\bibinfo  {journal} {Phys. Rev. Lett.}\ }\textbf {\bibinfo
  {volume} {62}},\ \bibinfo {pages} {2747} (\bibinfo {year}
  {1989})}\BibitemShut {NoStop}%
\bibitem [{\citenamefont {Grusdt}\ \emph {et~al.}(2019)\citenamefont {Grusdt},
  \citenamefont {Yao},\ and\ \citenamefont
  {Demler}}]{grusdt_topological_2019-1}%
  \BibitemOpen
  \bibfield  {author} {\bibinfo {author} {\bibfnamefont {F.}~\bibnamefont
  {Grusdt}}, \bibinfo {author} {\bibfnamefont {N.~Y.}\ \bibnamefont {Yao}},\
  and\ \bibinfo {author} {\bibfnamefont {E.~A.}\ \bibnamefont {Demler}},\
  }\href {https://doi.org/10.1103/PhysRevB.100.075126} {\bibfield  {journal}
  {\bibinfo  {journal} {Phys. Rev. B}\ }\textbf {\bibinfo {volume} {100}},\
  \bibinfo {pages} {075126} (\bibinfo {year} {2019})}\BibitemShut {NoStop}%
\bibitem [{Note5()}]{Note5}%
  \BibitemOpen
  \bibinfo {note} {For OBC systems, $P$ is well-behaved in the symmetry broken
  phase (in qualitative agreement with PBC), in the normal phase experiences a
  boundary-driven transition for moderate system sizes. This (finite-size
  effect) transition, which is due to edge states hybridizing with the bulk,
  disappears in the thermodynamic limit.}\BibitemShut {Stop}%
\bibitem [{Note6()}]{Note6}%
  \BibitemOpen
  \bibinfo {note} {The fractionalized value for $P$ is intrinsically due to CDW
  order.}\BibitemShut {Stop}%
\bibitem [{Note7()}]{Note7}%
  \BibitemOpen
  \bibinfo {note} {We have also performed these cycles explicitly via time
  evolution on small systems. The results are presented in the Supplementary
  Material~\cite {supplement}}\BibitemShut {NoStop}%
\bibitem [{Note8()}]{Note8}%
  \BibitemOpen
  \bibinfo {note} {Due to finite-size effects, we were forced to used small
  pinning field $|\Delta | = 10^{-2}$ throughout. In the thermodynamic limit
  however, $\Delta = 0$ would also work (the collective order can be induced by
  random fluctuations).}\BibitemShut {Stop}%
\bibitem [{Note9()}]{Note9}%
  \BibitemOpen
  \bibinfo {note} {These correlations also allow us to realize faster pumps, as
  explicitly illustrated in~\cite {supplement}}\BibitemShut {NoStop}%
\bibitem [{\citenamefont {McGinley}\ and\ \citenamefont
  {Cooper}(2018)}]{mcginley_topology_2018}%
  \BibitemOpen
  \bibfield  {author} {\bibinfo {author} {\bibfnamefont {M.}~\bibnamefont
  {McGinley}}\ and\ \bibinfo {author} {\bibfnamefont {N.~R.}\ \bibnamefont
  {Cooper}},\ }\href {https://doi.org/10.1103/PhysRevLett.121.090401}
  {\bibfield  {journal} {\bibinfo  {journal} {Phys. Rev. Lett.}\ }\textbf
  {\bibinfo {volume} {121}},\ \bibinfo {pages} {090401} (\bibinfo {year}
  {2018})}\BibitemShut {NoStop}%
\bibitem [{noa()}]{noauthor_itensor_nodate}%
  \BibitemOpen
  \href {https://itensor.org/index.html} {\bibinfo {title} {{ITensor}
  {Library}}}\BibitemShut {NoStop}%
\end{thebibliography}
\end{document}